\documentclass[%
 aip,
 jmp,%
 amsmath,amssymb,
 reprint,%
]{revtex4-1}

\usepackage{graphicx}% Include figure files
\graphicspath{{Figures/}}
\usepackage{dcolumn}% Align table columns on decimal point
\usepackage{bm}% bold math
\usepackage{subfigure}

\usepackage{algorithm}% http://ctan.org/pkg/algorithm
\usepackage{algpseudocode}% http://ctan.org/pkg/algorithmicx
\usepackage{amsmath,amssymb,amsfonts}
\usepackage{graphics}
\usepackage{color}

\begin{document}

%\preprint{AIP/123-QED}

\title[Data-driven deconvolution for large eddy simulations of Kraichnan turbulence]{Data-driven deconvolution for large eddy simulations of Kraichnan turbulence}% Force line breaks with \\
%\thanks{Footnote to title of article.}

\author{R. Maulik}
\author{O. San}%
 \email{osan@okstate.edu}
\affiliation{ 
School of Mechanical \& Aerospace Engineering, Oklahoma State University, Stillwater, Oklahoma - 74078, USA.
}%

\author{A. Rasheed}
 %\homepage{http://www.Second.institution.edu/~Charlie.Author.}
\affiliation{%
CSE Group, Mathematics and Cybernetics, SINTEF Digital, Trondheim, Norway.
}%

\author{P. Vedula}
 %\homepage{http://www.Second.institution.edu/~Charlie.Author.}
\affiliation{%
School of Aerospace \& Mechanical Engineering, The University of Oklahoma, Norman, Oklahoma - 73019, USA.
}%

\date{\today}% It is always \today, today,

\begin{abstract}
In this article, we demonstrate the use of artificial neural networks as optimal maps which are utilized for convolution and deconvolution of coarse-grained fields to account for sub-grid scale turbulence effects. We demonstrate that an effective eddy-viscosity is predicted by our purely data-driven large eddy simulation framework without explicit utilization of phenomenological arguments. In addition, our data-driven framework precludes the knowledge of true sub-grid stress information during the training phase due to its focus on estimating an effective filter and its inverse so that grid-resolved variables may be related to direct numerical simulation data statistically. The proposed predictive framework is also combined with a statistical truncation mechanism for ensuring numerical realizability in an explicit formulation. Through this we seek to unite structural and functional modeling strategies for modeling non-linear partial differential equations using reduced degrees of freedom. Both \emph{a priori} and \emph{a posteriori} results are shown for a two-dimensional decaying turbulence case in addition to a detailed description of validation and testing. A hyperparameter sensitivity study also shows that the proposed dual network framework simplifies learning complexity and is viable with exceedingly simple network architectures. Our findings indicate that the proposed framework approximates a robust and stable sub-grid closure which compares favorably to the Smagorinsky and Leith hypotheses for capturing the theoretical $k^{-3}$ scaling in Kraichnan turbulence.
\end{abstract}

% \pacs{Valid PACS appear here}% PACS, the Physics and Astronomy
%                              % Classification Scheme.
%Use showkeys class option if keyword
\maketitle

% \begin{quotation}
% The ``lead paragraph'' is encapsulated with the \LaTeX\ 
% \verb+quotation+ environment and is formatted as a single paragraph before the first section heading. 
% (The \verb+quotation+ environment reverts to its usual meaning after the first sectioning command.) 
% Note that numbered references are allowed in the lead paragraph.
% %
% The lead paragraph will only be found in an article being prepared for the journal \textit{Chaos}.
% \end{quotation}

\section{Introduction}

Over the past decade, advances in data collection and increasing access to computational resources have led to a revolution in the use of data-driven techniques for the solution of intractable inverse problems \cite{ARCS,ARBM,guest2018deep,duraisamy2018turbulence}. One such problem is that of turbulence, the multiscale nature of which causes infeasible computational demands even for the most simple systems. This behavior is shared by all non-linear partial differential equations and necessitates the utilization of multiple modeling approximations for tractable compute times. One such modeling approach is that of large eddy simulation (LES) \cite{sagaut2006large}, which attempts to simulate the evolution of lower wavenumber modes of turbulence while the effects of higher wavenumber modes are modeled by an algebraic or differential equation. The procedure of modeling the finer scales is often denoted a \emph{closure} due to the lack of knowledge about higher-order wavenumber interactions in the coarse-grained flow  \cite{berselli2006mathematics} and remains a critical component of accurate computational modeling for many applications \cite{hickel2014subgrid,yu2016dynamic,zhou2018structural}. From an LES point of view, the closure problem arises due to the fact that low-pass spatial filtering (due to coarse-graining and discrete numerical approximations) does not commute with the non-linear term.

Within the context of the Navier-Stokes equations, it is generally accepted that the finer scales are dissipative at the Kolmogorov length scales \cite{kolmogorov1941local} and therefore, most turbulence models seek to specify a sub-grid viscosity which mimics the dissipative behavior of the unsupported frequencies \cite{frisch1995turbulence}. Most sub-grid models can be traced back to the seminal work of Smagorinsky \cite{smagorinsky1963general}, where a model was proposed based on the concepts of an effective eddy viscosity determined by an \emph{a priori} specified mixing length and a $k^{-5/3}$ scaling recovery for the kinetic energy content in the wavenumber domain. Similar hypotheses have also been used for two-dimensional turbulence \cite{leith1968diffusion} (often utilized as a test-bed for geophysical scenarios, for instance see works by Pearson \textit{et al.}\cite{pearson2018log,pearson2017evaluation}), for approximating the $k^{-3}$ cascade in two-dimensional turbulence and generally have their roots in dimensional analysis related to the cascade of enstrophy. The two aforementioned models may be classified as functional due to the phenomenological nature of their deployment and represent the bulk of LES related turbulence models used in practical deployments.

In contrast, the structural approach to turbulence modeling utilizes no explicit specification of an eddy-viscosity and relies on an estimation of the low-pass spatial filtering nature of coarse-graining. With this approximate knowledge of the filter, arguments for scale-similarity \cite{bardina1980improved,layton2003simple} or approximate-deconvolution (AD) \cite{stolz1999approximate} are utilized to reconstruct the true non-linear term. In case of scale-similarity, the non-linear interactions of flow components are estimated by utilizing a forward filtering operation to the grid-resolved variables, while in AD an inverse filter is estimated using iterative resubstitutions. However, structural techniques are limited due to the fact that they approximately recover sub-filter stresses alone and are not dissipative enough due to the neglect of sub-grid considerations. Therefore, they require the specification of an additional (usually functional) sub-grid model or the specification of a finer resolution where sub-grid terms are negligible \cite{germano2015similarity}. Further information about turbulence models and whether they may be classified as functional or structural may be found in Saugaut's excellent text \cite{sagaut2006large}.

A common thread that connects both functional and structural models is the \emph{a priori} specification of a model coefficient or a characteristic filter width or ratio. Consequently, the choice of such parameters become crucial in the \emph{a posteriori} performance of the deployed model. Crucially, literature has consistently shown that the choice of these coefficients are not single-valued, particularly for off-nominal flow situations. One may refer to discussions by Galperin and Orszag \cite{galperin1993large} and Canuto and Cheng \cite{canuto1997determination} for examples for the effect of varying eddy viscosity. The effect of characteristic filter widths and the order of deconvolution has also been explored by San \textit{et al.}\cite{san2015posteriori} and by Schneiderbauer and Saeedipour\cite{schneiderbauer2018approximate}. With this contextual background, in this study, we introduce a hybrid modeling (physics-informed machine learning) methodology for determining sub-grid models without any phenomenological assumptions (in the spirit of structural models) but with sub-grid capture ability. This is accomplished by the use of artificial neural networks (ANNs) to establish data-driven maps between \emph{a priori} convolved and deconvolved fields but without the use of any explicit filter.

In recent times, data-driven techniques have become extremely popular for the spatio-temporal modeling of dynamical systems \cite{schmidt2009distilling,bright2013compressive,xiao2015non,brunton2016discovering,schaeffer2017learning,raissi2017machine,mohan2018deep,raissi2018hidden,rudy2018deep,san2018neural,wan2018data,kim2018deep,muravleva2018application,jin2018prediction}. With respect to turbulence, some widely used strategies for inference include symbolic regression \cite{weatheritt2016novel,weatheritt2017development,weatheritt2017hybrid}, where functional model-forms for RANS deployments were generated through optimization against high-fidelity data. Ma \textit{et al.}\cite{ma2015using} utilized compressive-sensing based machine learning for closure of multiphase system. Gautier \textit{et al.}\cite{gautier2015closed} utilized a genetic algorithm was utilized for regression tasks in a close-loop separation control deployment of a turbulent mixing layer. Other techniques incorporating Bayesian ideologies have also been used, for instance by Xiao \textit{et al.}\cite{xiao2016quantifying} where an iterative ensemble Kalman method was used to assimilate prior data for quantifying model form uncertainty in RANS models. In Wang \textit{et al.}\cite{wang2017physics,wang2017comprehensive} and Wu \textit{et al.}\cite{wu2018data}, random-forest regressors were utilized for RANS turbulence-modeling given DNS data. In Singh and Duraisamy \cite{singh2016using} and Singh \textit{et al.}\cite{singh2017machine}, an ANN was utilized to predict a non-dimensional correction factor in the Spalart-Allmaras turbulence model through a field-inversion process. The field-inversion process was utilized to develop optimal \emph{a priori} estimates for the correction factor from experimental data. Bypassing functional formulations of a turbulence model (a focus of this study) was also studied from the RANS point of view by Tracey \textit{et al.} \cite{tracey2015machine}. Ling and Templeton \cite{ling2015evaluation} utilized support vector machines, decision trees and random forest regressors for identifying regions of high RANS uncertainty. A deep-learning framework where Reynolds-stresses would be predicted in an invariant subspace was developed by Ling \textit{et al.} \cite{ling2016reynolds}. The reader is directed to a recent review by Duraisamy \textit{et al.}\cite{duraisamy2018turbulence}, for an excellent review of turbulence modeling using data-driven ideas.

As shown above, the use of machine learning ideologies and in particular ANNs has generated significant interest in the turbulence modeling community. This is motivated by the fact that a multilayered artificial neural network may be optimally trained to universally approximate any non-linear function \cite{hornik1989multilayer}. Greater accessibility to data and the GPU revolution has also motivated the development of advanced ANN architectures for constrained learning and improved physical interpretability. Within the context of LES (and associated with the scope of this paper) there are several investigations into sub-grid modeling using data-driven techniques. In one of the first studies of the feasibility of mapping to unresolved stresses using grid resolved variables by learning from high-fidelity data, Sarghini \textit{et al.}\cite{sarghini2003neural} utilized ANNs for estimating Smagorinsky model-form coefficients within a mixed sub-grid model for a turbulent channel flow. This may be considered similar to the field-inversion procedure describe previously. ANNs were also used for wall-modeling by Milano and Koumotsakos \cite{milano2002neural} where it was used to reconstruct the near wall field and compared to standard proper-orthogonal-decomposition techniques. An alternative to ANNs for sub-grid predictions was proposed by King \textit{et al.}\cite{king2016autonomic} where \emph{a priori} optimization was utilized to minimize the $L^2$-error between true and modeled sub-grid quantities in a least-squares sense using a parameter-free Volterra series. Maulik and San \cite{maulik2017neural} utilized an extreme-learning-machine (a variant of a single-layered ANN) to obtain maps between low-pass spatially filtered and deconvolved variables in an \emph{a priori} sense. This had implications for the use of ANNs for turbulence modeling without model-form specification. A more in-depth investigation has recently been undertaken by Fukami \textit{et al.}\cite{fukami2018super} where convolutional ANNs are utilized for reconstructing downsampled snapshots of turbulence. Gamahara and Hattori \cite{gamahara2017searching}, utilized ANNs for identifying correlations with grid-resolved quantities for an indirect method of model-form identification in turbulent channel flow. The study by Vollant \textit{et al.} \cite{vollant2017subgrid} utilized ANNs in conjuction with optimal estimator theory to obtain functional forms for sub-grid stresses. In Beck \textit{et al.}\cite{beck2018neural}, a variety of neural network architectures such as convolutional and recurrent neural networks are studied for predicting closure terms for decaying homogeneous isotropic turbulence. A least-squares based truncation is specified for stable deployments of their model-free closures. Model-free turbulence closures are also specified by Maulik \textit{et al.}\cite{maulik2019subgrid}, where sub-grid scale stresses are learned directly from DNS data and deployed in \emph{a posteriori} through a truncation for numerical stability. King \textit{et al.}\cite{king2018deep} studied generative-adversarial networks and the LAT-NET \cite{hennigh2017lat} for \emph{a priori} recovery of statistics such as the intermittency of turbulent fluctuations and spectral scaling. A detailed discussion of the potential benefits and challenges of deep learning for turbulence (and fluid dynamics in general) may be found in the article by Kutz \cite{kutz2017deep}.

While a large majority of the LES-based frameworks presented above utilize a least-squares error minimization technique for constructing maps to sub-grid stresses \emph{directly}, this work represents a physics-informed implementation of sub-grid source terms through the learning of convolutional and deconvolution maps between grid-resolved and unresolved fields. In other words, our framework is able to reproduce, approximately, a map related to the convolution associated with insufficient grid-support in LES implementations of the Navier-Stokes equations as well as its inverse. These optimal maps are obtained by supervised learning from subsampled direct numerical simulation (DNS) data and are deployed in an \emph{a posteriori} fashion for the LES of two-dimensional turbulence. In this manner, we unite the advantages of functional and structural modeling of turbulence in addition to precluding the use of any phenomenological arguments. Through this, we also aim to achieve a harmonious combination of first-principles based physics as well data-driven mechanisms for high accuracy. A hybrid formulation leveraging our knowledge of governing equations and augmenting these with machine learning represents a great opportunity for obtaining optimal LES closures for multiscale physics simulations \cite{langford1999optimal,moser2009theoretically,labryer2015framework,king2016autonomic,pathak2018hybrid}.

Therefore, this investigation represents an advancement of the concepts proposed by the authors previously \cite{maulik2017neural}, where solely the deconvolutional ability of artificial neural networks was investigated in an \emph{a priori} sense for sub-filter stresses. The adaptations proposed in our current study are targeted towards recovering the sub-grid component of the coarse-grained LES computation. In addition, we not only address the issue of \emph{a priori} sub-grid recovery with our proposed closure, but also demonstrate its robustness in \emph{a posteriori} deployment with associated numerical challenges. While the two-dimensional turbulence case is utilized for a proof-of-concept as well as for its geophysical implications where improved closure development is still sought extensively, our generalized framework may easily be scaled up to multiple dimensional non-linear partial differential equations. Our results indicate that the proposed framework provides for a robust sub-grid model with a dynamically computed effective eddy-viscosity within the structural modeling ideology.

\section{Turbulence modeling equations}

We proceed with the introduction of our framework by outlining the governing equations for two-dimensional turbulence. These are given by the Navier-Stokes equations in the vorticity-streamfunction formulation. In place of a primitive variable formulation, our decaying turbulence problem is solved for using the temporal evolution of the following non-dimensionalized and coupled system of equations,
\begin{align}
\label{Eq1a}
\begin{split}
\frac{\partial \omega}{\partial t} + J(\omega,\psi) &= \frac{1}{Re} \nabla^2 \omega, \\
\nabla^2 \psi &= -\omega,
\end{split}
\end{align}
where the velocity vector components may be recovered as
\begin{align}
\label{Eq1b}
\begin{split}
\frac{\partial \psi}{\partial y} &= u \\
\frac{\partial \psi}{\partial x} &= -v.
\end{split}
\end{align}
The computational necessities of coarse-graining result in a grid-filtered system of equations
\begin{align}
\label{Eq2}
\begin{split}
\frac{\partial \overline{\omega}}{\partial t} + J(\overline{\omega},\overline{\psi}) &= \frac{1}{Re} \nabla^2 \overline{\omega} + \Pi, \\
\nabla^2 \overline{\psi} &= -\overline{\omega},
\end{split}
\end{align}
where overbarred quantities imply grid-resolved variables. A resulting unclosed term is obtained, ideally represented as
\begin{align}
\label{Eq3}
\Pi = J(\overline{\omega},\overline{\psi}) - \overline{J(\omega,\psi)}.
\end{align}
The second term on the right-hand side of the above equation represents the primary target of approximation for the structural modeling mechanism. In contrast, the functional modeling procedure is to represent $\Pi$ as an effective eddy-viscosity multiplied by Laplacian of the vorticity. In this study, we shall utilize a data-driven paradigm for approximating
\begin{align}
\label{Eq4}
\overline{J(\omega,\psi)} \approx \widetilde{J(\omega^*,\psi^*)},
\end{align}
where asterisked quantities are those obtained by data-driven deconvolution and the tilde represents data-driven convolution. This procedure is similar to the AD mechanism which requires an \emph{a priori} low-pass spatial filter specification. Note that the proposed methodology effectively aims to approximate the operations of Fourier cut-off filtering and its inverse which is the primary reason why it blends the distinction between sub-filter and sub-grid recovery. The former is a potential limitation of the AD mechanism in its current implementation. Our approximate sub-grid model is thus given by
\begin{align}
\label{Eq4b}
\tilde{\Pi} = J(\bar{\omega},\bar{\psi})-\widetilde{J(\omega^*,\psi^*)}.
\end{align}
For the purpose of comparison we also introduce the Smagorinsky and Leith models which utilize algebraic eddy-viscosities for sub-grid stress calculation given by
\begin{align}
\label{Eq5}
\Pi_e = \nabla . \left(\nu_e \nabla \bar{\omega}\right),
\end{align}
where for the Smagorinsky model we have
\begin{align}
\label{Eq6}
\nu_e = (C_s \delta)^2 |\bar{S}|,
\end{align}
and the Leith hypothesis states
\begin{align}
\label{Eq7}
\nu_e = (C_l \delta)^3 |\nabla \bar{\omega}|.
\end{align}

Note that $|\bar{S}| = \sqrt{2 S_{ij} S_{ij}}$ and $|\nabla \bar{\omega}|$ correspond to two commonly used kernels for eddy-viscosity approximations. Here, $\delta$ is generally assumed to be the characteristic mixing length taken to be the grid size. The online performance of our proposed framework shall be compared to these simple, but robust closures. We remark here that the standard procedure for closure in the vorticity-streamfunction formulation (relevant to two-dimensional simulations) is based on sub-grid vorticity source term modeling but our generalized procedure may be extended to the primitive variable approach as a source term in the Navier-Stokes momentum equations. For the convenience of the reader we also tabulate some of the notation that will be widely used in the rest of this article in Table \ref{Table1}. We note that the variables outlined in this table are all defined on a coarse(i.e, LES) grid. Details regarding the preparation of the data for our machine learning methods shall be outlined in subsequent sections.

\begin{table}[H]
\label{Table1}
\small
\centering
\resizebox{\columnwidth}{!}{%
\begin{tabular}{|c|c|} \hline
 Notation      & Category  \\ \hline
 $\bar{a}$     & Grid filtered (i.e, Fourier cut-off filtered) from DNS   \\ \hline
 $a^c$         & Comb filtered (i.e, sub-sampled) from DNS   \\ \hline
 $a^*$         & Data-driven deconvolved variable  \\ \hline
 $\tilde{a}$   & Data-driven convolved variable  \\   \hline
\end{tabular}
}
\caption{A summary of filter and deconvolutional notation}
\end{table}

\section{Data-driven convolution and deconvolution}

The ANN, also known as a multilayered perceptron, consists of a set of linear or nonlinear mathematical operations on an input space vector to establish a map to an output space. Other than the input and output spaces, a ANN may also contain multiple hidden layers (denoted so due to the obscure mathematical significance of the matrix operations occurring here). Each of these layers is an intermediate vector in a multi-step transformation which is acted on by biasing and activation before the next set of matrix operations. Biasing refers to an addition of a constant vector to the incident vector at each layer, on its way to a transformed output. The process of activation refers to an elementwise functional modification of the incident vector to generally introduce nonlinearity into the eventual map. In contrast, no activation (also referred to as a linear activation), results in the incident vector being acted on solely by biasing. Note that each component of an intermediate vector corresponds to a unit cell also known as the neuron. The learning in this investigation is \emph{supervised} implying labeled data used for informing the optimal map between inputs and outputs. Mathematically, if our input vector $\textbf{p}$ resides in a $P$-dimensional space and our desired output $\textbf{q}$ resides in a $Q$-dimensional space, the ANN establishes a map $\mathbb{M}$ as follows:
\begin{align}
\label{eq8}
\mathbb{M} : \{ p_1, p_2, \hdots, p_P\} \in \mathbb{R}^P \rightarrow \{ q_1, q_2, \hdots, q_Q\} \in \mathbb{R}^Q.
\end{align}

In this study, we utilize two maps which relate to convolution and deconvolution of fields with grid-resolved and sub-grid components respectively. We must caution the reader here that the maps are not assumed to transform between isomorphic spaces (considered a limitation of structural AD \cite{guermond2004mathematical,germano2015similarity}). This allows for the estimation of sub-grid loss due to coarse-graining the degrees of freedom in an LES deployment. In equation form, our optimal map $\mathbb{M}_1$ relates coarse-grained field stencils to their grid-filtered (i.e., Fourier cut-off filtered) counterparts and is given by
\begin{align}
\label{eq9}
\begin{gathered}
\mathbb{M}_1 : \{ \omega_{i,j}^c, \omega_{i,j+1}^c, \omega_{i,j-1}^c \hdots, \omega_{i-1,j-1}^c \in \mathbb{R}^{9} \rightarrow \{ \tilde{\omega} \} \in \mathbb{R}^1.
\end{gathered}
\end{align}
where $\tilde{\omega}$ represents an approximation for $\bar{\omega}$.

Our second map, relates grid-filtered field stencils to their coarse-grained counterparts given by
\begin{align}
\label{eq10}
\begin{gathered}
\mathbb{M}_2 : \{ \bar{\omega}_{i,j}, \bar{\omega}_{i,j+1}, \bar{\omega}_{i,j-1} \hdots, \bar{\omega}_{i-1,j-1} \in \mathbb{R}^{9} \rightarrow \{ \omega^{*} \} \in \mathbb{R}^1.
\end{gathered}
\end{align}
where $\omega^{*}$ represents an approximation for $\omega^c$. Note that both maps are trained for optimal prediction using normalized inputs. Our normalization (approximately) rescales our data to zero mean and unit variance by using grid-resolved variable quantities. Therefore, both inputs and outputs to maps are normalized by quantities available dynamically and the deployment of the network does not require \emph{a priori} storage of training parameters. For instance, the normalization of $\bar{\omega}$ may be obtained by
\begin{align}
\label{Eq11a}
\bar{\omega}^n = \frac{\bar{\omega}-\mu(\bar{\omega})}{\sigma(\bar{\omega})},
\end{align}
where $\mu(a)$ and $\sigma(a)$ refer to the mean and variance of a field variable $a$. Similarly the normalization of $\omega^{*}$ is given by
\begin{align}
\label{Eq11b}
\omega^{*^{n}} = \frac{\omega^{*}-\mu(\bar{\omega})}{\sigma(\bar{\omega})}.
\end{align}
In this manner, no \emph{a priori} training coefficients may be recorded. In essence, we emphasize that all normalization is carried out to ensure the mean of grid-resolved quantities is zero and that the standard deviation of these quantities is unity. Trained maps using this normalization technique may thus be used for the convolution or deconvolution of any coarse-grained variable.

Note that both maps are trained for optimal prediction using normalized inputs. Our normalization (approximately) rescales our data to zero mean and unit variance by using grid-resolved variable quantities. Therefore, both inputs and outputs to maps are normalized by quantities available dynamically. A key facet of our proposed methodology is that our trained maps are obtained only from vorticity data even though they need deployment for the deconvolution of the streamfunction as well as the convolution of the Jacobian. Successful sub-grid information recovery (described in the results section) shows that this data-independence in training can be related to a true learning of the filtering and deconvolution characteristics between coarse and fine grids.

\begin{figure}
\centering
\caption{A schematic of our data-driven mapping for convolution and deconvolution. Two separate ANNs are utilized for projection to and from deconvolved variable space.}
\includegraphics[width=\columnwidth,trim=0cm 1cm 0cm 0cm,clip]{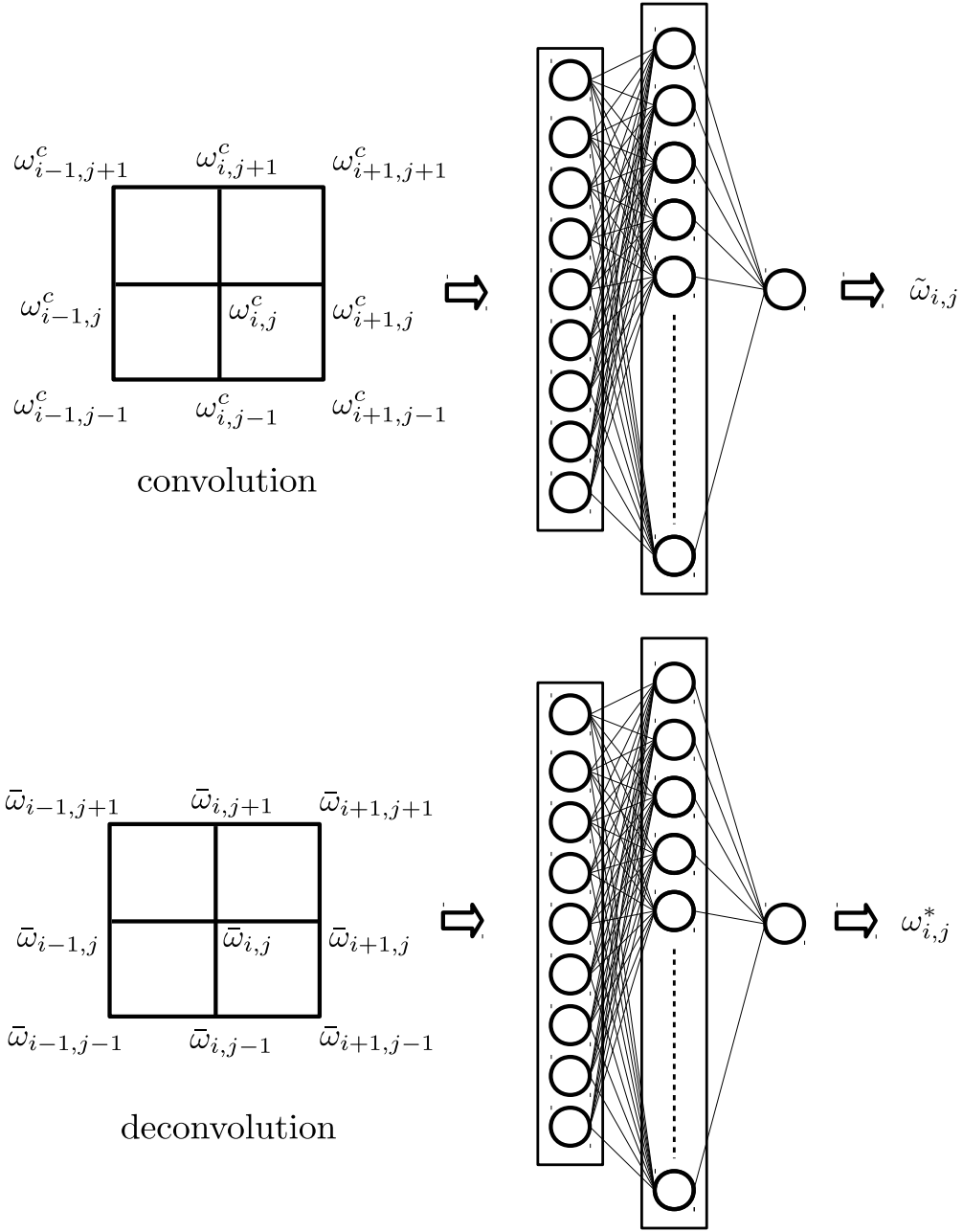}
\label{Fig0}
\end{figure}

The pseudocode for a deployment of our optimally trained maps is shown in Algorithm \ref{Algo1} where it can be seen that each time step (or sub-step) of an explicit flow evolution requires the specification of a data-driven approximation to the true Jacobian $J\overline{(\omega,\psi)}$. In subsequent sections, we shall comment on the final \emph{a posteriori} constraining for ensuring numerical realizability. Figure \ref{Fig0} visually outlines the two networks deployed in this study.

\begin{algorithm}[H]
  \caption{Proposed framework deployment}
  \label{Algo1}
  \begin{algorithmic}[1]
        \State Given trained maps $\mathbb{M}_1 \textnormal{ and } \mathbb{M}_2$

        \State Given $\overline{\omega} \textnormal{ and } \overline{\psi}$

        \State Normalize $\overline{\omega} \textnormal{ and } \overline{\psi}$ to get $\overline{\omega}^n \textnormal{ and } \overline{\psi}^n$ respectively

        \State Use $\mathbb{M}_2$ to obtain deconvolved variables $\omega^{n^*} \textnormal{ and }\psi^{n^*}$

        \State Rescale to physical domain to get $\omega^{*} \textnormal{ and } \psi^{*}$

        \State Calculate estimated coarse-grid Jacobian $J(\omega^*,\psi^*)$

        \State Normalize Jacobian $J(\omega^*,\psi^*)$ to get $J(\omega^*,\psi^*)^n$

        \State Use $\mathbb{M}_1$ to obtain convolved variables $\widetilde{J(\omega^*,\psi^*)^n}$

        \State Rescale $\widetilde{J(\omega^*,\psi^*)^n}$ to physical domain to get $\widetilde{J(\omega^*,\psi^*)}$

        \State Deploy turbulence model $\tilde{\Pi} = J(\bar{\omega},\bar{\psi}) - \widetilde{J(\omega^*,\psi^*)}$ subject to post-processing for numerical stability given by Equation \ref{Eq12}

  \end{algorithmic}
\end{algorithm}

\begin{figure}
\centering
\caption{A schematic of our biasing and activation at each hidden layer neuron. Assuming five inputs from previous layer.}
\includegraphics[width=\columnwidth,trim=0cm 14.5cm 0cm 5cm,clip]{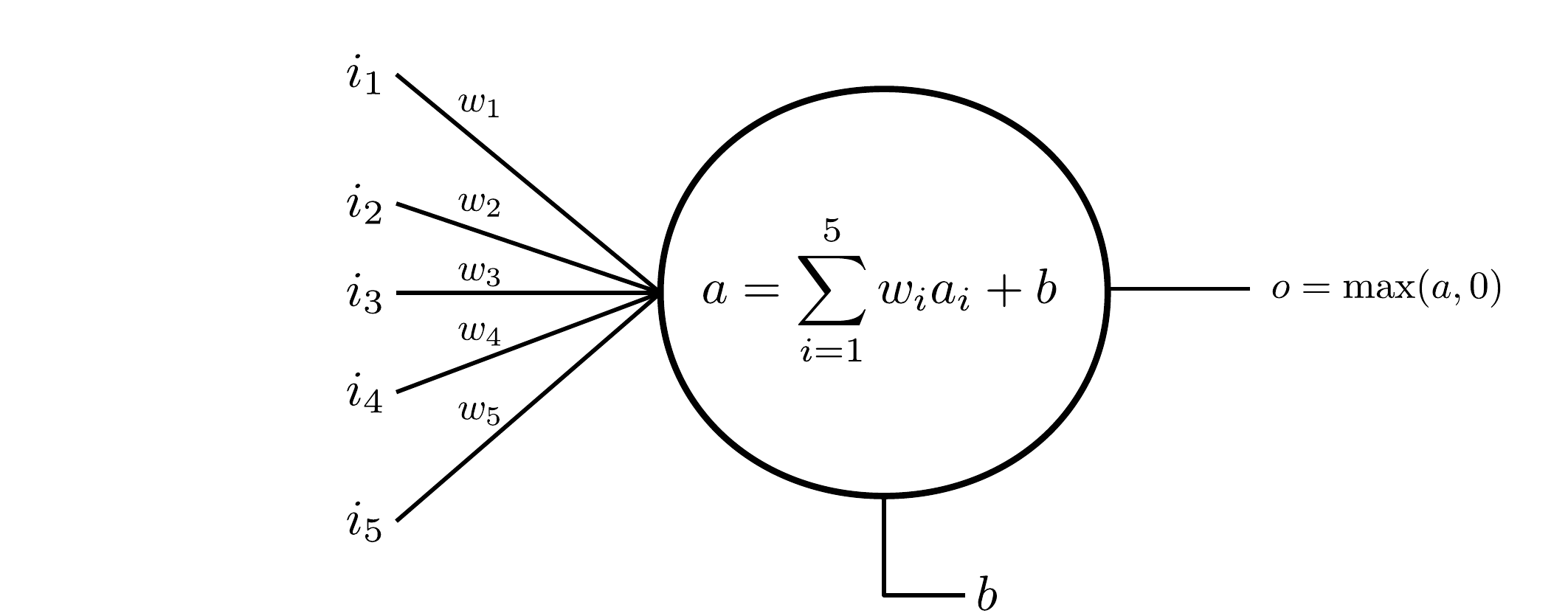}
\label{Fig1}
\end{figure}

As evident, implementation of the proposed framework requires multiple convolutional and deconvolutional passes over the grid-resolved variables and therefore we refer to this framework, from henceforth, as the data-driven convolutional and deconvolutional closure (DCD). Both our networks utilize one hidden layer along with the input and output layers. This hidden and output layers have a bias vector associated with it. For faster training, we utilize rectified linear activation functions (ReLU) for our hidden layer and a linear activation function for the output layer. Note that input data is not activated as it enters the network. Our hidden layer utilizes 100 unit cells (i.e., neurons) which are acted on by the ReLU transformation and biasing. The process of bias and activation at each neuron is displayed in Figure \ref{Fig1} and every neuron is fully connected to its previous layer (i.e, with incident inputs from all neurons from the previous layer). In subsequent sections, we outline a sensitivity study of our proposed ideology for varying architecture depths where it is proven that one-layered networks suffice for this particular problem.

\section{Training and \emph{a priori} validation}

For the purpose of generating our optimal maps discussed in the previous section, we utilize two supervised learnings with sets of labeled inputs and outputs obtained from direct numerical simulation (DNS) data for two-dimensional Kraichnan turbulence. We have utilized a second-order accurate energy-conserving Arakawa scheme for the nonlinear Jacobian and second-order accurate finite-difference discretization schemes for the Laplacian of the vorticity. The Poisson update is performed using a spectrally-accurate solver and the time-integration is performed by a third-order accurate TVD Runge-Kutta explicit method. Further details on the problem setup and the implementation of an energy and enstrophy conserving numerical method can be found by the authors' previous studies \cite{san2012high,maulik2017stable}. Our grid-resolved variables (i.e., $\bar{\omega}$) are generated by a Fourier cut-off filter so as to truncate the fully-resolved DNS fields (obtained at $2048^2$ degrees-of-freedom) to coarse-grained grid level (i.e., given by $256^2$ degrees-of-freedom). Our subsampled variables (i.e., $\omega^c$) are obtained by a comb filtering procedure where every eighth data point is retained.

% Therefore, these two types of filtering, outlined in a generalized schematic for one-dimension given by Figure \ref{Fig2}, are utilized to generate input-output vorticity pairs for the process of training our ANN maps.

% \begin{figure}
% \centering
% \caption{A schematic of coarse-graining procedures: (a) Fourier cut-off grid filtering approach to obtain $\bar{a}$ on $x \in 256$ resolution from high fidelity data $a$ on $x \in 2048$ resolution and (b) comb filtering approach for obtaining sub-sampling of data of $a^c$ on $x \in 256$ resolution from high fidelity data $a$. Note that it is only shown here for a one-dimensional data array and its extension to a multidimensional data array is straightforward.}
% \includegraphics[width=\columnwidth]{Figure_2.pdf}
% \label{Fig2}
% \end{figure}

We also emphasize on the fact that, while the DNS data generated multiple time snapshots of flow evolution, data was harvested from times $t=0,1,2,3$ and $4$ for the purpose of training and validation. This represents a stringent subsampling of the total available data for map optimization. Our DNS utilized an explicit formulation with a constant timestep of 0.0001 implying potential generation of 40000 snapshots out of which only 4 were selected at regular intervals for data harvesting. This represents a 0.01\% utilization of total potential data during training which is particularly challenging for this unsteady problem. The generation of data sets at the coarse grained level is outlined in Algorithm \ref{Algo2}.

We also note that the Reynolds number chosen for generating the training and validation data sets is given by $Re=32000$ while deployment is tested for a higher Reynolds number of $64000$ for both \emph{a priori} and \emph{a posteriori} assessment. We remind the reader here, map training is performed solely on the vorticity field despite the fact that trained maps are to be utilized for vorticity, streamfunction and the Jacobian. For this reason, all our inputs are normalized to ensure zero mean and unit variance while our outputs are normalized in a similar fashion but to slightly different mean and variance i.e.,
\begin{align}
\label{Eq11}
a^n &= \frac{a - \mu(\bar{a})}{\sigma(\bar{a})},
\end{align}
where $a$ may either be grid-resolved or deconvolved quantities. In essence, we emphasize that all normalization is carried out to ensure the mean of grid-resolved quantities is zero and that the standard deviation of these quantites is unity. The aforementioned normalized quantities are then used as input-output pairs for the two different networks as discussed previously. The generation of data sets at the coarse grained level is outlined in algorithm \ref{Algo2}.

\begin{algorithm}[H]
  \caption{Data harvesting from DNS}
  \label{Algo2}
  \begin{algorithmic}[1]
        \State Obtain DNS data for vorticity $\omega^{DNS}$ at $N^2=2048^2$
        \State Comb filter to obtain $\omega^c$ from $\omega^{DNS}$ by sub-sampling every eighth point
        \State Grid filter to obtain $\bar{\omega}$ from $\omega^{DNS}$

        \State Normalize $\bar{\omega}$ to $\bar{\omega}^n$ using Equations \ref{Eq11a}

        \State Normalize $\omega^c$ to $\omega^{c^n}$ using Equation \ref{Eq11b}

        \State $\omega^{c^n}$ and $\bar{\omega}^n$ are input and output pairs respectively for map $\mathbb{M}_1$ optimization, where we assume true output $\tilde{\omega}^n \approx \bar{\omega}^n$ according to Equation \ref{Eq4}

        \State $\bar{\omega}^n$ and $\omega^{c^n}$ are input and output pairs respectively for map $\mathbb{M}_2$ optimization, where we assume true output $\omega^{*^n} \approx \omega^{c^n}$
  \end{algorithmic}
\end{algorithm}

Two-thirds of the total dataset generated for optimization is utilized for training and the rest is utilized for test assessment. Here, training refers to the use of data for loss calculation (which in this study is a classical mean-squared-error) and backpropagation for parameter update. The test data, however, is utilized to record the performance of the trained network on data it was not exposed to during training. Similar behavior in training and test losses would imply a well-formulated learning problem. The final ANN (obtained post-training) would be selected according to the best loss on the test data after a desired number of iterations which for this study was fixed at 50. The choice for a low number of iterations was observed by Pearson correlation values reaching 0.99 for both training and test data sets. We also note that the error-minimization in the training of the ANN utilized the Adam optimizer \cite{kingma2014adam} implemented in the open-source neural network training platform TensorFlow. We remark that while the networks may have learned the target maps from the data they are provided for training and test, validation would require an \emph{a posteriori} examination as detailed in the following section. We note here that data preprocessing as well as architectural modifications (for instance network depth, number of neurons and activation types) need further investigation for improved generalization. 

\begin{figure}
\centering
\includegraphics[width=\columnwidth]{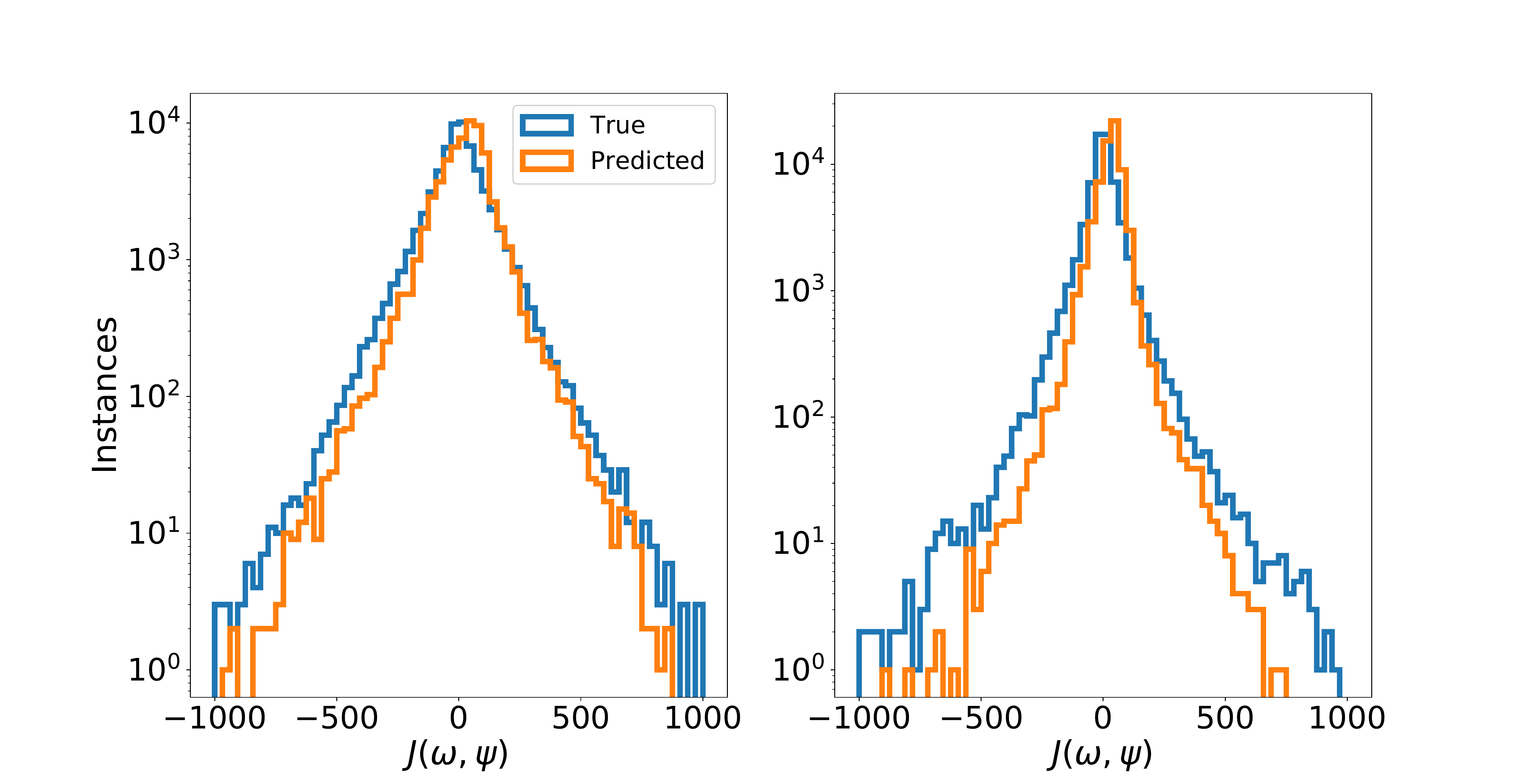}
\caption{The prediction ability of the use of both forward and inverse maps in the calculation of the approximate underlying Jacobian $\widetilde{J(\omega^{*},\psi^{*})}$ for $Re=32000$ (left) and $Re=64000$ (right). The true Jacobian $\overline{J(\omega,\psi)}$ is also shown.}
\label{Fig3}
\end{figure}

We first outline an \emph{a priori} study for the proposed framework where the optimal maps are utilized for predicting probability distributions for the true Jacobian i.e., $J(\overline{\omega,\psi})$. A pseudocode for the computation of this true Jacobian is outlined in Algorithm \ref{Algo3}. In other words, we assess the turbulence model for a one snapshot prediction. This study is carried out for one of our data snapshots $t=2$ but for both in and out-of-training data sets. We remark that the maps have previously been exposed to vorticity data from $Re=32000$ only and our out-of-training data set is given by a similar flow scenario but at higher Reynolds number given by $Re=64000$. One can thus make the argument for some transfer of learning between similar flow classes but with slight difference in physics. The performance of the framework is shown in Figure \ref{Fig3} where the framework predicts the density functions of the true Jacobian accurately for both sets of data. We also note that this study solely utilized a mean-squared-error minimization for the target variables without any physics-based regularization. A future study involving loss-functions devised with intuition from the Navier-Stokes equations would potentially aid in preserving invariance and symmetry properties between grid-resolved and deconvolved space. In addition, while the localized stencil based sampling for map deployments proposed here is amenable to deployment in structured grids, extension to arbitrary meshes would require the use of interpolation or graph convolutional kernels for unstructured information injection into the learning architecture.

\begin{algorithm}[H]
  \caption{True Jacobian $\overline{J(\omega,\psi)}$ from DNS}
  \label{Algo3}
  \begin{algorithmic}[1]
        \State Obtain DNS data for vorticity $\omega^{DNS}$ and streamfunction $\psi^{DNS}$ at $N^2=2048^2$
        \State Calculate Jacobian on DNS grid i.e., $J(\omega^{DNS},\psi^{DNS})$
        \State Apply grid filter to $J(\omega^{DNS},\psi^{DNS})$ in order to obtain $\overline{J(\omega,\psi)}$ at $N^2=256^2$.
  \end{algorithmic}
\end{algorithm}

\section{\emph{A posteriori} testing}

\begin{figure}
\centering
\includegraphics[width=\columnwidth]{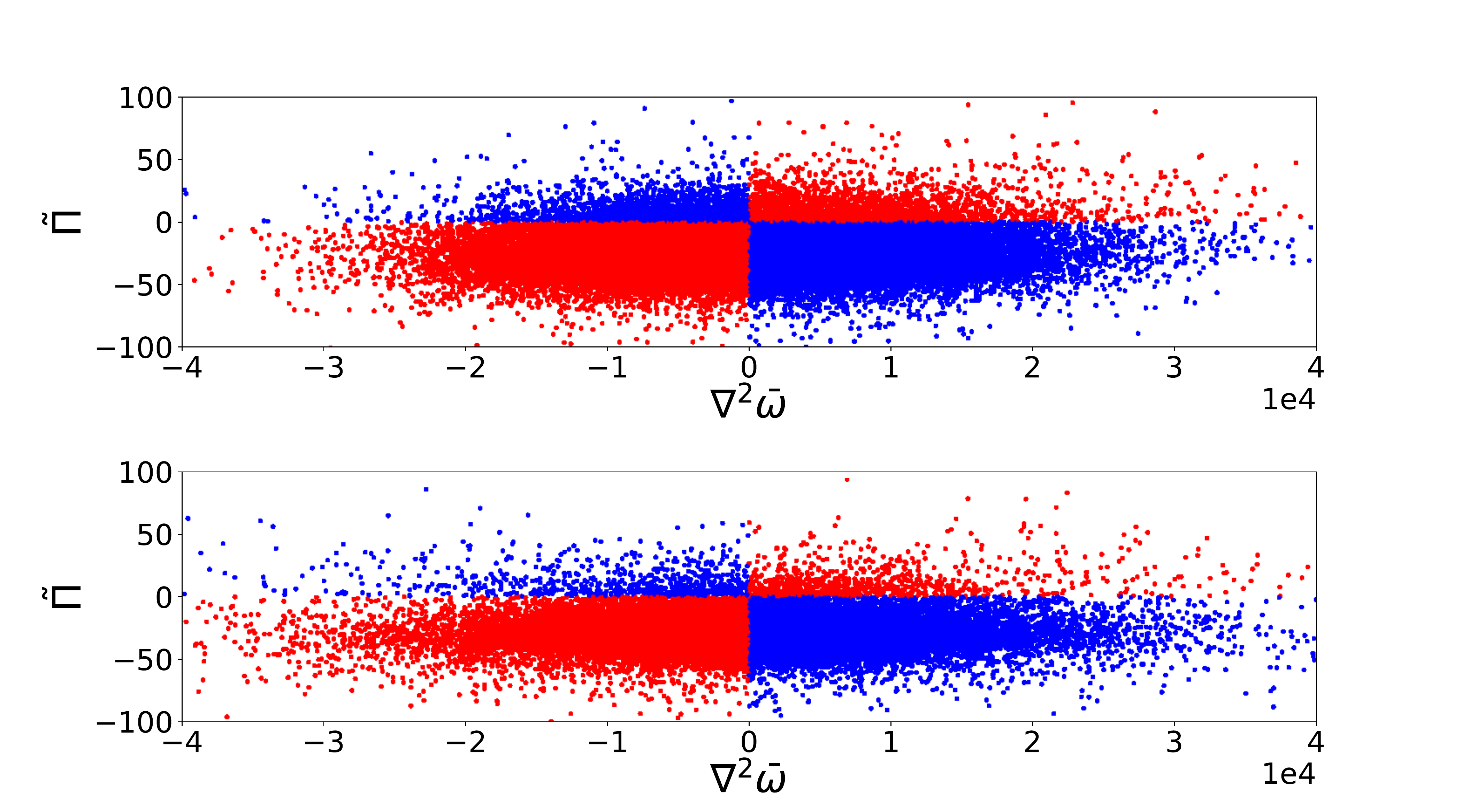}
\caption{A visual assessment of the truncation of our numerical post-processing during deployment given by Equation \ref{Eq12}. Blue points indicate truncated deployment for ensuring no negative viscosity and numerical stability. A-priori predictions for $Re=32000$ (top) and $Re=64000$ (bottom) shown. }
\label{Fig4}
\end{figure}

The ultimate test of any data-driven closure model is in an \emph{a posteriori} framework with subsequent assessment for the said model's ability to preserve coherent structures and scaling laws. While the authors have undertaken \emph{a priori} studies with promising results for data-driven ideologies for LES \cite{maulik2017neural}, the results of the following section are unique in that they represent a model-free turbulence model computation in temporally and spatially dynamic fashion. This test setup is particulary challenging due to the neglected effects of numerics in the \emph{a priori} training and testing. In the following we utilize angle-averaged kinetic energy spectra to assess the ability of the proposed framework to preserve integral and inertial range statistics. Theoretical comparisons with Kraichnan turbulence \cite{kraichnan1967inertial} and the expected $k^{-3}$ cascade are also provided.  In brief, we mention that the numerical implementation of the conservation laws are through second-order discretizations for all spatial quantities (with a kinetic-energy conserving Arakawa discretization for the calculation of the nonlinear Jacobian). A third-order total-variation-diminishing Runge-Kutta method is utilized for the vorticity evolution and a spectrally-accurate Poisson solver is utilized for updating streamfunction values from the vorticity. Our proposed framework is deployed pointwise for estimating $\tilde{\Pi}$ at each explicit time-step until the final time of $t=4$ is reached. The robustness of the network to the effects of numerics is thus examined. For the purpose of numerical stability we ensure the following condition before deploying our framework
\begin{align}
\label{Eq12}
\Pi =
\begin{cases}
\tilde{\Pi},& \text{if  } (\nabla^2 \bar{\omega}) (\tilde{\Pi}) > 0\\
    0,              & \text{otherwise.}
\end{cases}
\end{align}
where the truncation explicitly ensures no negative numerical viscosities due to the deployment of the sub-grid model. We remind the reader that the Smagorinsky and Leith hypotheses explicitly specify positive eddy-viscosities that are obtained by absolute value quantities as given in Equations \ref{Eq6} and \ref{Eq7}. An \emph{a priori} visual quantification of the truncation is shown in Figure \ref{Fig4} where quantities in the first and third quadrants are retained predictions and the others are discarded. A similar behavior is seen for both $Re=32000$ and $Re=64000$ data. This image also highlights the challenges of translating \emph{a priori} conclusions to \emph{a posteriori} implementations due to the requirement of numerical stability.

\begin{figure}
\centering
\includegraphics[width=\columnwidth]{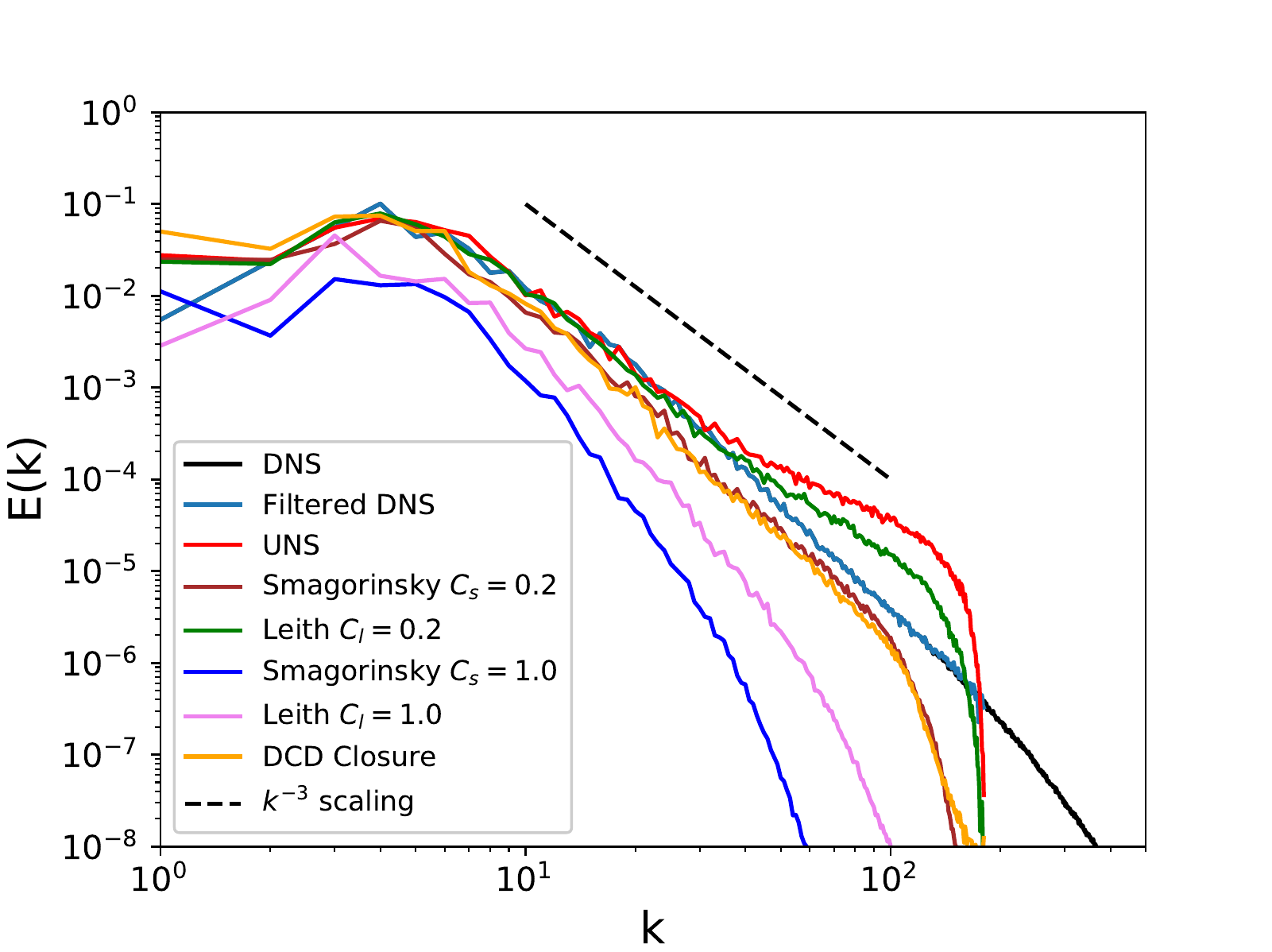}
\caption{The \emph{a posteriori} performance of proposed framework for $Re=32000$ in terms of energy spectra. At each step of sub-grid stress calculation, both forward and inverse maps are used for convolution and deconvolution in the estimation of the true underlying Jacobian.}
\label{Fig5}
\end{figure}

\begin{figure}
\centering
\centering
\includegraphics[width=\columnwidth]{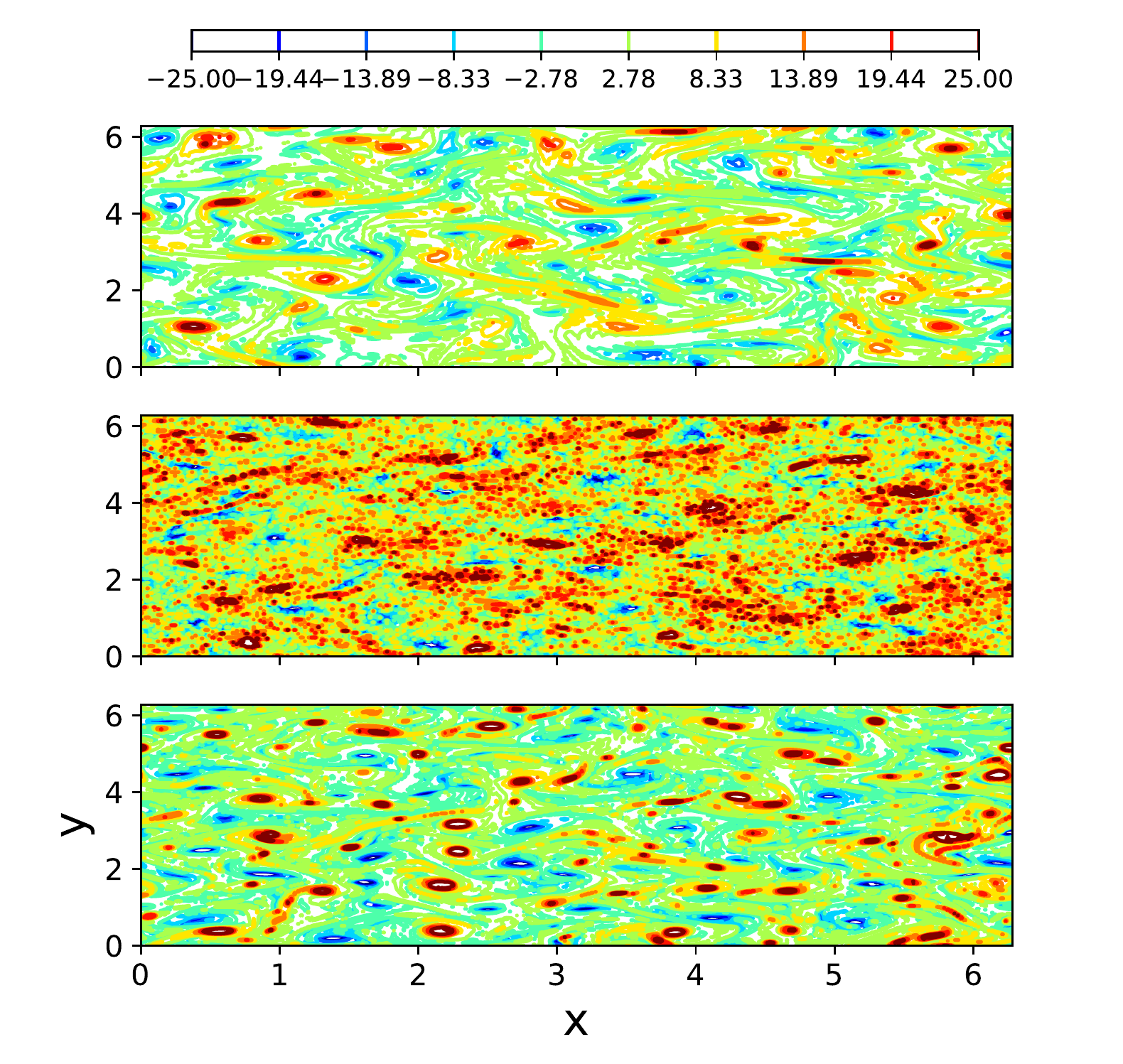}
\caption{Visual quantification of the \emph{a posteriori} performance of proposed framework for $Re=32000$ with stabilized (top), under-resolved (middle) and filtered DNS contours (bottom) for vorticity.}
\label{Fig6}
\end{figure}

\begin{figure}
\centering
\includegraphics[width=\columnwidth]{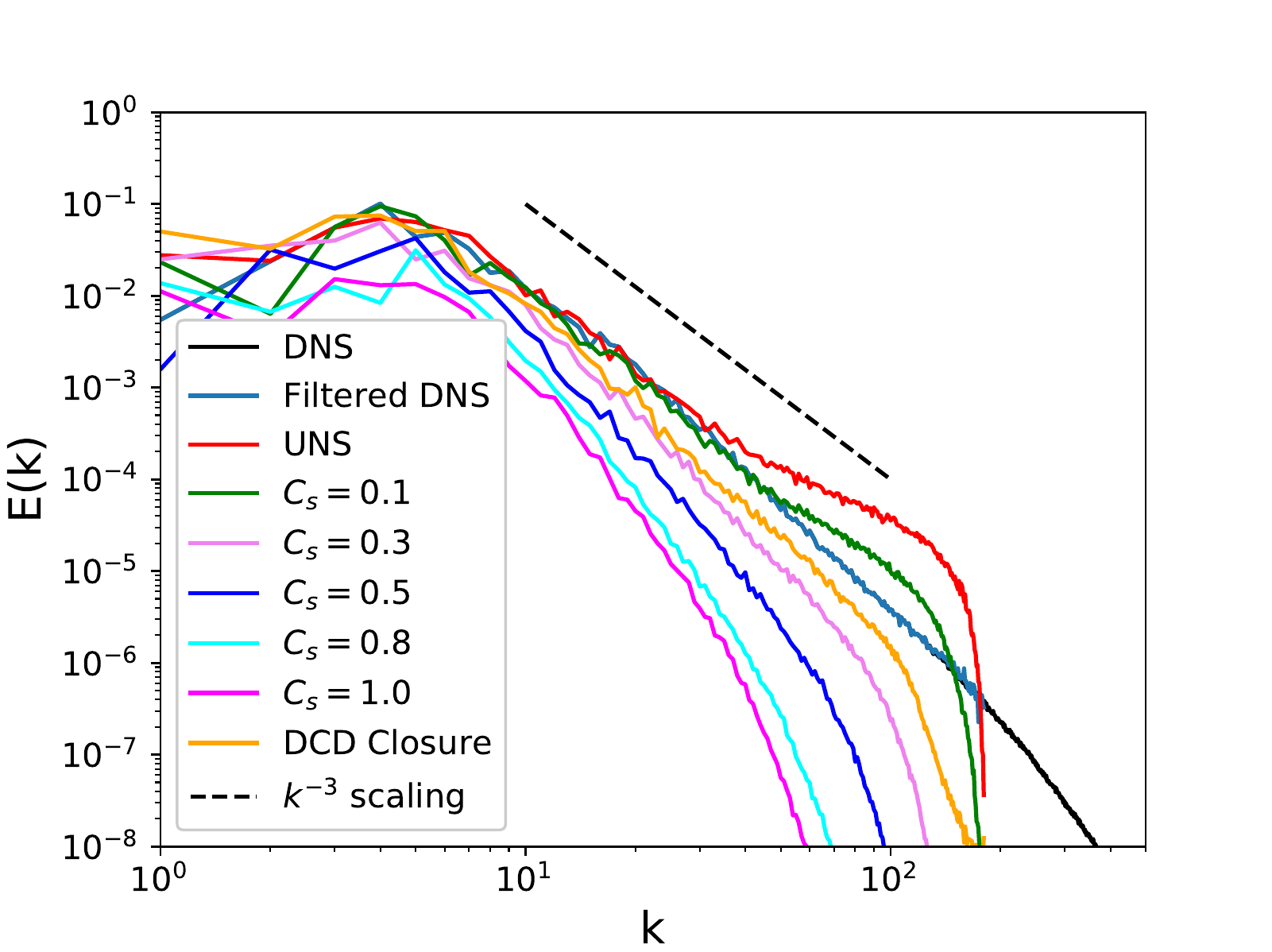}
\caption{Performance comparison of proposed framework with co-efficient dependant Smagorinsky model. One can observe that higher $C_s$ values lead to over-dissipative models.}
\label{Fig7}
\end{figure}

\begin{figure}
\centering
\includegraphics[width=\columnwidth]{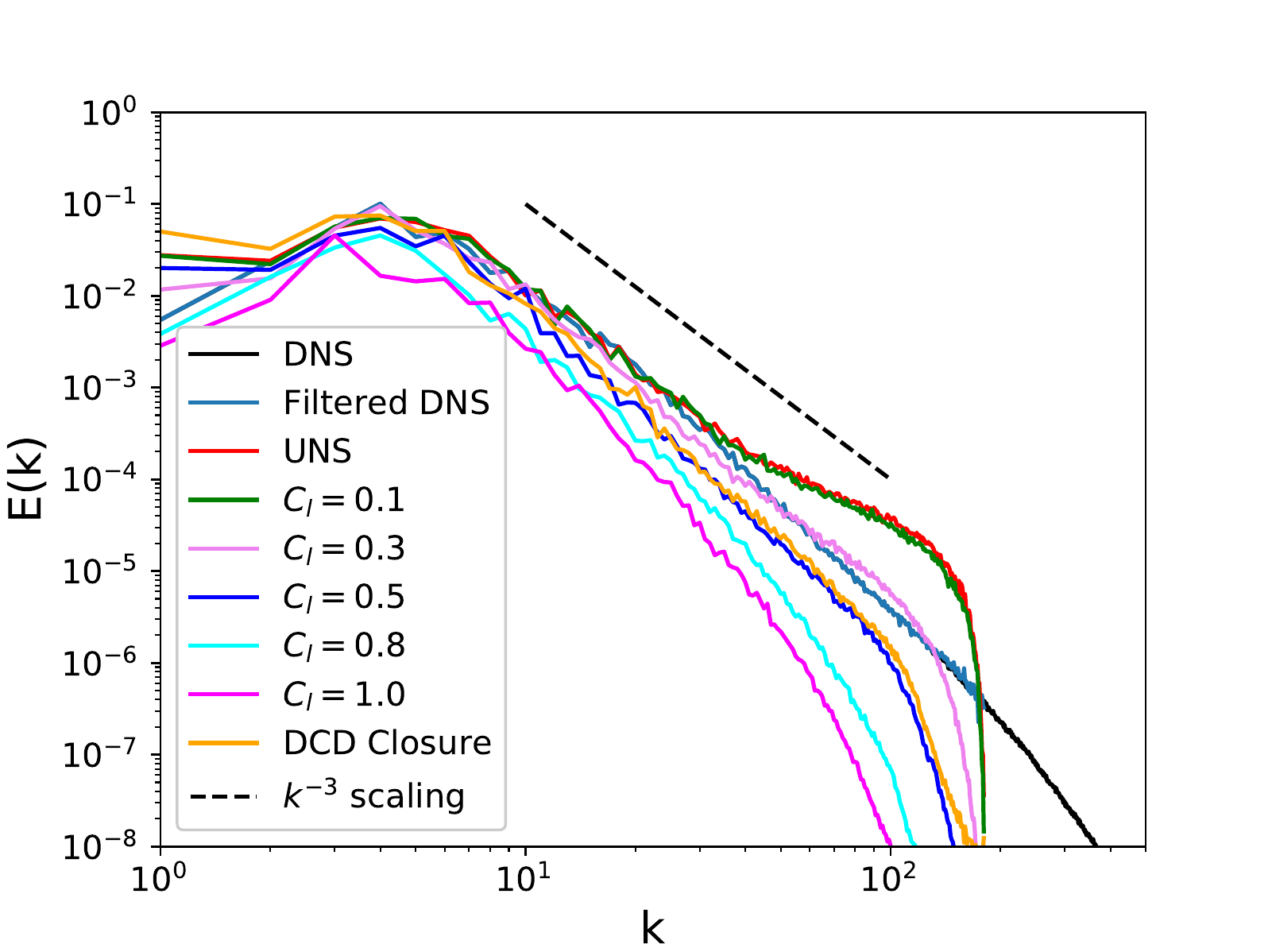}
\caption{Performance comparison of proposed framework with co-efficient dependant Leith model. One can observe that higher $C_l$ values lead to over-dissipative models.}
\label{Fig8}
\end{figure}

Figure \ref{Fig5} displays the statistical fidelity of coarse-grained simulations obtained with the deployment of the proposed framework for $Re=32000$. Stable realizations of the vorticity field are generated due to the combination of our training and post-processing. For the purpose of comparison, we also include coarse-grained no-model simulations, i.e., unresolved numerical simulations (UNS) which demonstrate an expected accumulation of noise at grid cut-off wavenumbers. DNS spectra are also provided showing agreement with the $k^{-3}$ theoretical scaling expected for two-dimensional turbulence. Our proposed framework is effective at stabilizing the coarse-grained flow by estimating the effect of sub-grid quantities and preserving trends with regards to the inertial range scaling. Figure \ref{Fig6} visually quantifies the effect of the stabilization imparted by the proposed framework. The reader may observe that the proposed framework recovers an excellent scaling behavior. This is similar to the performance obtained by deploying the Smagorinsky model at $C_s=0.2$, a widely utilized parameteric choice obtained through prior numerical experimentation. The Leith performance at $C_l=0.2$ is slightly under-dissipative. The reader may notice that an arbitrary choice of $C_s=C_l=1.0$ leads to overdissipative performance of the eddy-viscosity closures. Our data-driven framework is thus more resistant to unnecessary dissipation. Note that choice of a higher eddy viscosity coefficient for two-dimensional turbulence has been detailed in previous literature \cite{cushman2011introduction}. Another quantification of the perfomance of the DCD closure is described in Figures \ref{Fig7} and \ref{Fig8} which juxtapose the varying performance of these parameter-dependant eddy-viscosity hypothesis (i.e., Smagorinsky and Leith respectively) to the proposed data-driven approach.  One can observe that an optimal selection of parameters (after \emph{a posteriori} examination) given by $C_l = 0.5$ for the Leith model recreates the performance of the proposed framework well as well. This implies that the proposed framework has learned a similar dissipative nature through \emph{a priori} optimization of a filter and its inverse. Indeed, the application of the Smagorinsky model to various engineering and geophysical flow problems has revealed that the constant is not single-valued and varies depending on resolution and flow characteristics \cite{galperin1993large,canuto1997determination,vorobev2008smagorinsky} with higher values specifically for geophysical flows. In comparison, the proposed framework has embedded the adaptive nature of dissipation into its map which is a promising outcome. Before proceeding, we note that default parameteric choices for the Smagorinsky and Leith models are given by $C_s=C_l=0.2$.

\begin{figure}
\centering
\includegraphics[width=\columnwidth]{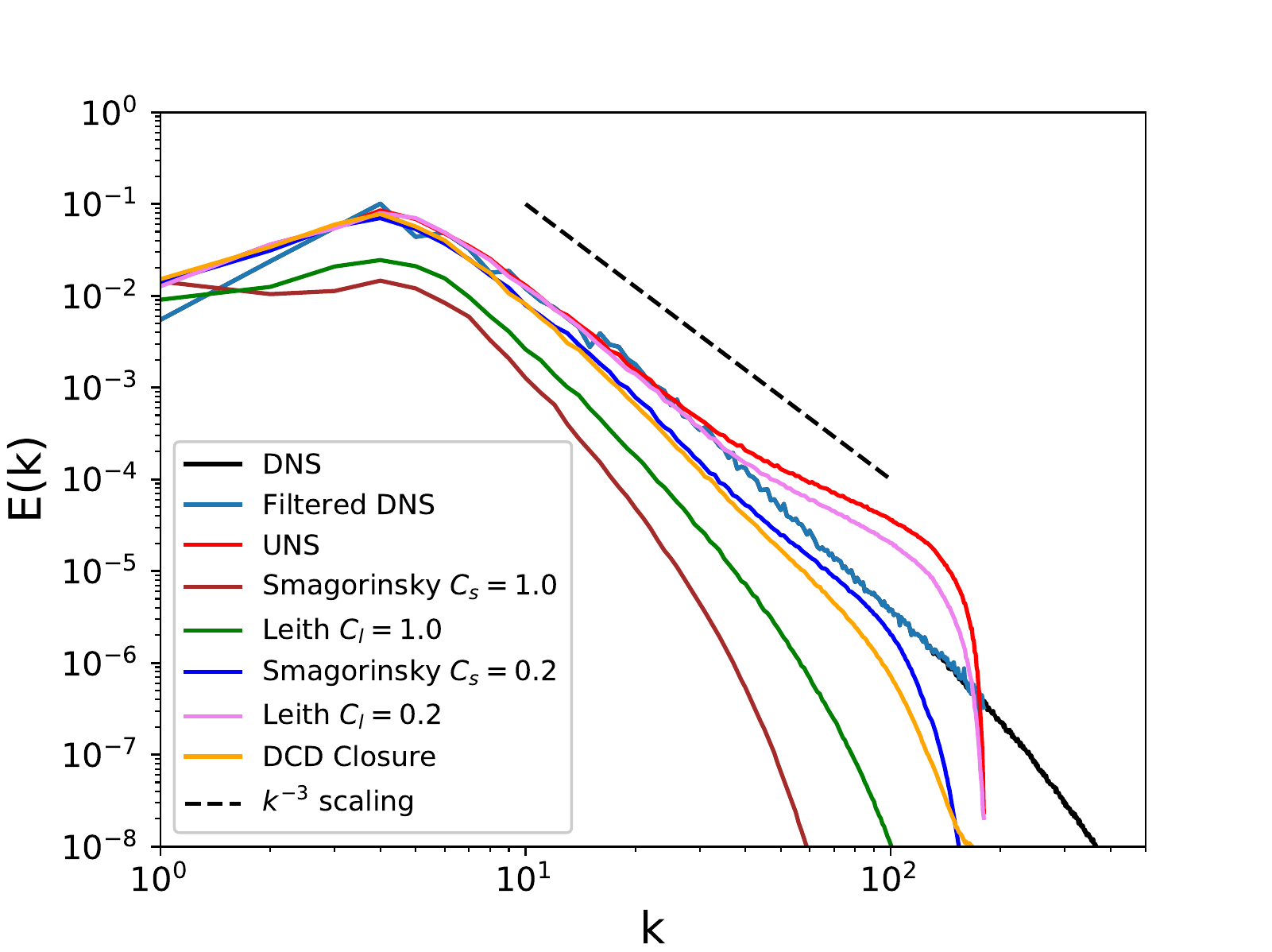}
\caption{Ensemble-averaged \emph{a posteriori} performance of proposed framework for $Re=32000$ in terms of energy spectra. This determines the generalizability of proposed framework.}
\label{Fig9}
\end{figure}

\begin{figure}
\centering
\includegraphics[width=\columnwidth]{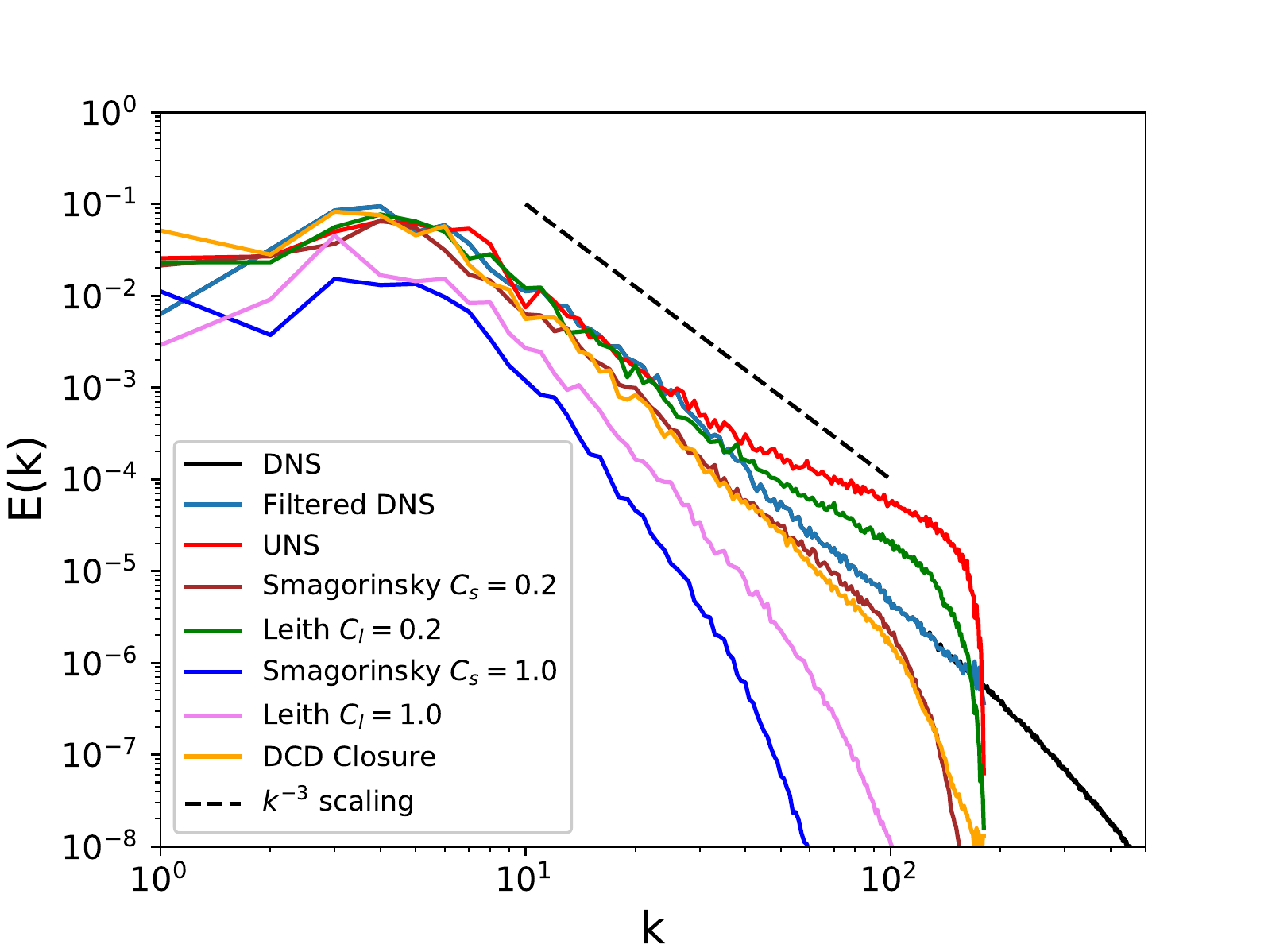}
\caption{The \emph{a posteriori} performance of proposed framework for $Re=64000$ in terms of energy spectra. Training data limited to $Re=32000$ only.}
\label{Fig10}
\end{figure}

\begin{figure}
\centering
\centering
\includegraphics[width=\columnwidth]{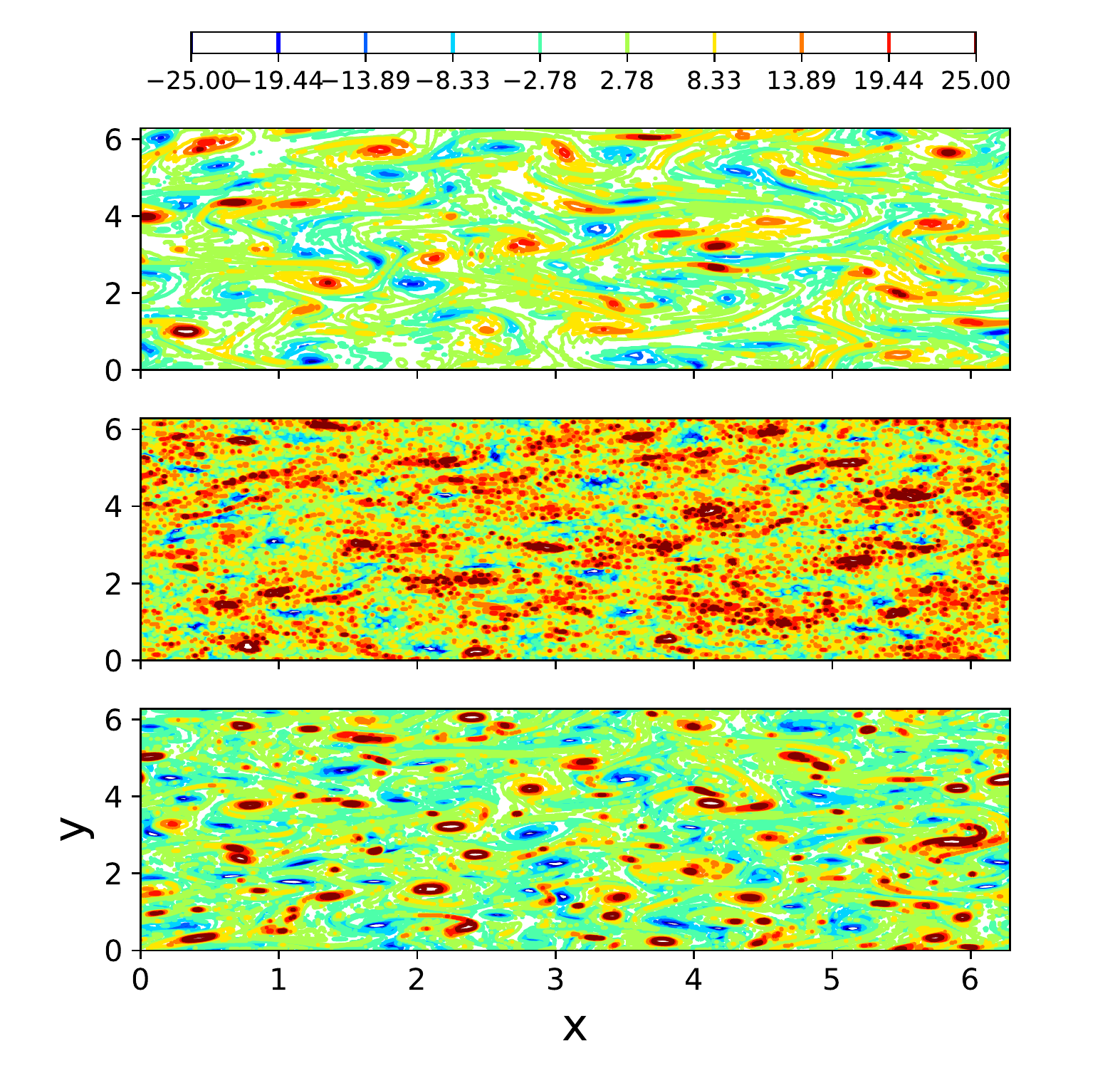}
\caption{Visual quantification of the \emph{a posteriori} performance of proposed framework for $Re=64000$ with stabilized (top), under-resolved (middle) and filtered DNS contours (bottom) for vorticity. Note: Training only with $Re=32000$ data.}
\label{Fig11}
\end{figure}

For ensuring that the training is sufficiently generalized for this particular problem, we establish a suite of testing for the predictive performance and the numerical stability of our proposed framework. We first perform multiple forward simulations using the deployment of our proposed closure by utilizing a different random seed in the random-number generation required for the initial conditions at $Re=32000$ \cite{maulik2017stable}. This is to ensure that there is no data memorization by our maps. We choose 24 random initial conditions and ensemble-average their kinetic energy spectra at the completion of the LES for our model as well as the Smagorinsky, Leith and no-model (i.e., UNS) coarse-grid runs. We have also included ensemble results from Smagorinsky and Leith deployments at higher values of $C_s=C_l=1.0$ to describe the loss of fidelity at the lower wavenumbers in case of incorrect parameter specification. The resultant spectra are shown in Figure \ref{Fig9} where one can ascertain that the prediction quality of our framework remains identical regardless of varying initial conditions. This is promising as it validates our hypothesis that it is the smaller scales which are primarily affected by the proposed closure.  We also demonstrate the utility of our learned map on an \emph{a posteriori} simulation for $Re=64000$ data where similar trends are recovered as seen in statistical comparisons (Figure \ref{Fig10}) and qualitative behavior (Figure \ref{Fig11}). This also demonstrates an additional stringent validation of the data-driven model for ensuring generalization.

\begin{figure}
\centering
\includegraphics[width=\columnwidth]{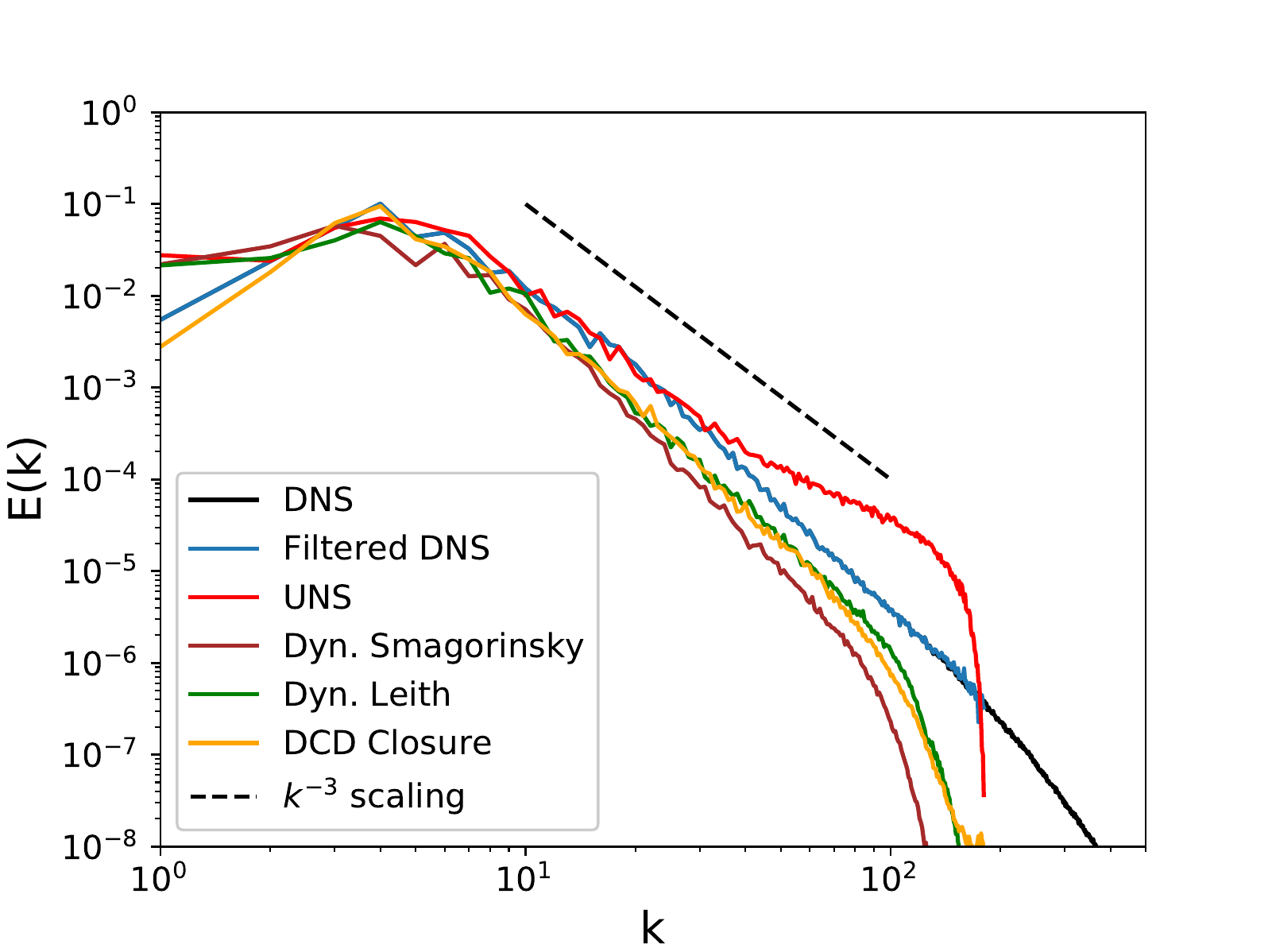}
\caption{A comparison of the proposed framework with the Dynamic Smagorinsky and Dynamic Leith models for $Re=32000$. One can see an optimal solution being obtained by the data-driven formulation in a similar manner.}
\label{FigRev3}
\end{figure}

\begin{figure}
\centering
\includegraphics[width=\columnwidth]{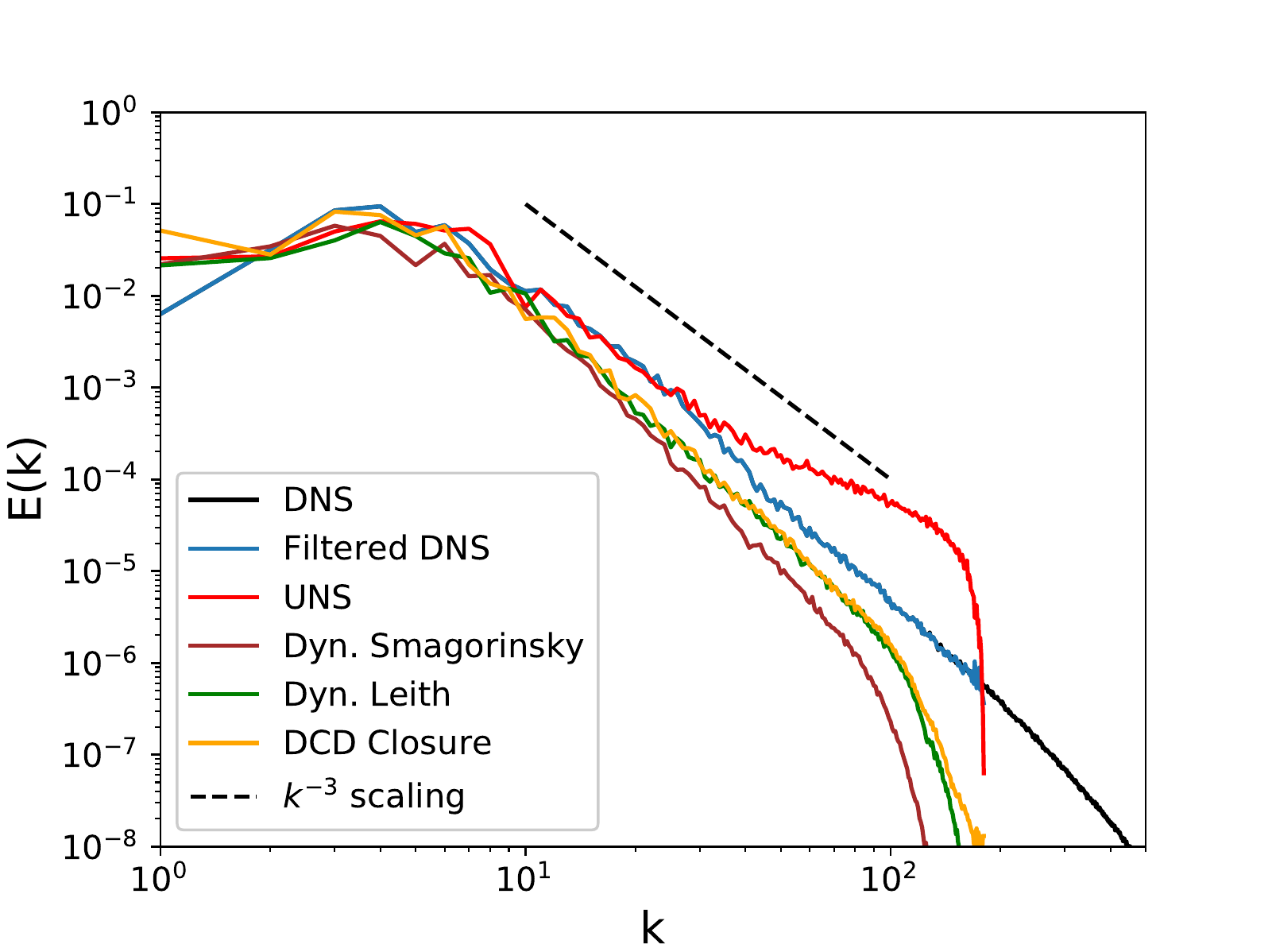}
\caption{A comparison of the proposed framework with the Dynamic Smagorinsky and Dynamic Leith models for $Re=64000$. One can see an optimal solution being obtained by the data-driven formulation in a similar manner. Training data limited to $Re=32000$ only.}
\label{FigRev4}
\end{figure}

We also seek to compare the performance of the proposed framework against the dynamic formulation of the Smagorinsky and Leith models \cite{germano1991dynamic} modified for the vorticity and streamfunction formulation as described by San and Maulik \cite{maulik2017stable} where a least-squares optimization problem is solved at two scales of resolution for an optimal value of the Smagorinsky and Leith coefficients calculated in a dynamic fashion defining a test filter. We note that even the dynamic formulation requires the specification of an \emph{a priori} characteristic filter-width ratio (i.e., a ratio between test and grid filters), $\kappa$, which affects \emph{a posteriori} results. In this comparison, we have utilized a filter-width ratio of $\kappa=2$ with the use of an explicit trapezoidal filter.  The results of this comparison with our framework are shown for Reynolds numbers of $Re=32000$ and $Re=64000$ in Figures \ref{FigRev3} and \ref{FigRev4} respectively. One can observe that the performance of the dynamic implementations of our eddy-viscosity hypotheses are recreated in a qualitative fashion. Our model may thus be assumed to be both data-driven and dynamic in nature.

In terms of computational cost, we remark that the proposed framework adds a considerable computational expenditure (\emph{a posteriori} simulations led to 4 times the computational cost of the dynamic formulation) in the serial formulation. However, scalable deployments of the proposed framework in distributed environments are a subject of ongoing investigation for reducing this cost. While the data-driven framework promises more accuracy through exposure to multiple sources of turbulence data, its scalable deployment remains an important open question for successful integration into modern computational fluid dynamics solvers.

\section{Sensitivity study}

We investigate the robustness of our framework by ensuring that an optimal number of hidden layers or neurons have been utilized through an \emph{a posteriori} sensitivity study where a varying number of layers and neurons are tested for spectral scaling recovery. By keeping the default network architecture as a one layer, 100 neuron network, we investigate the effect of reduction or increase in neurons as well the effect of the number of hidden layers. We note that our studies are performed for $Re=64000$ as an additional cross-validation.

\begin{figure}
\centering
\includegraphics[width=\columnwidth]{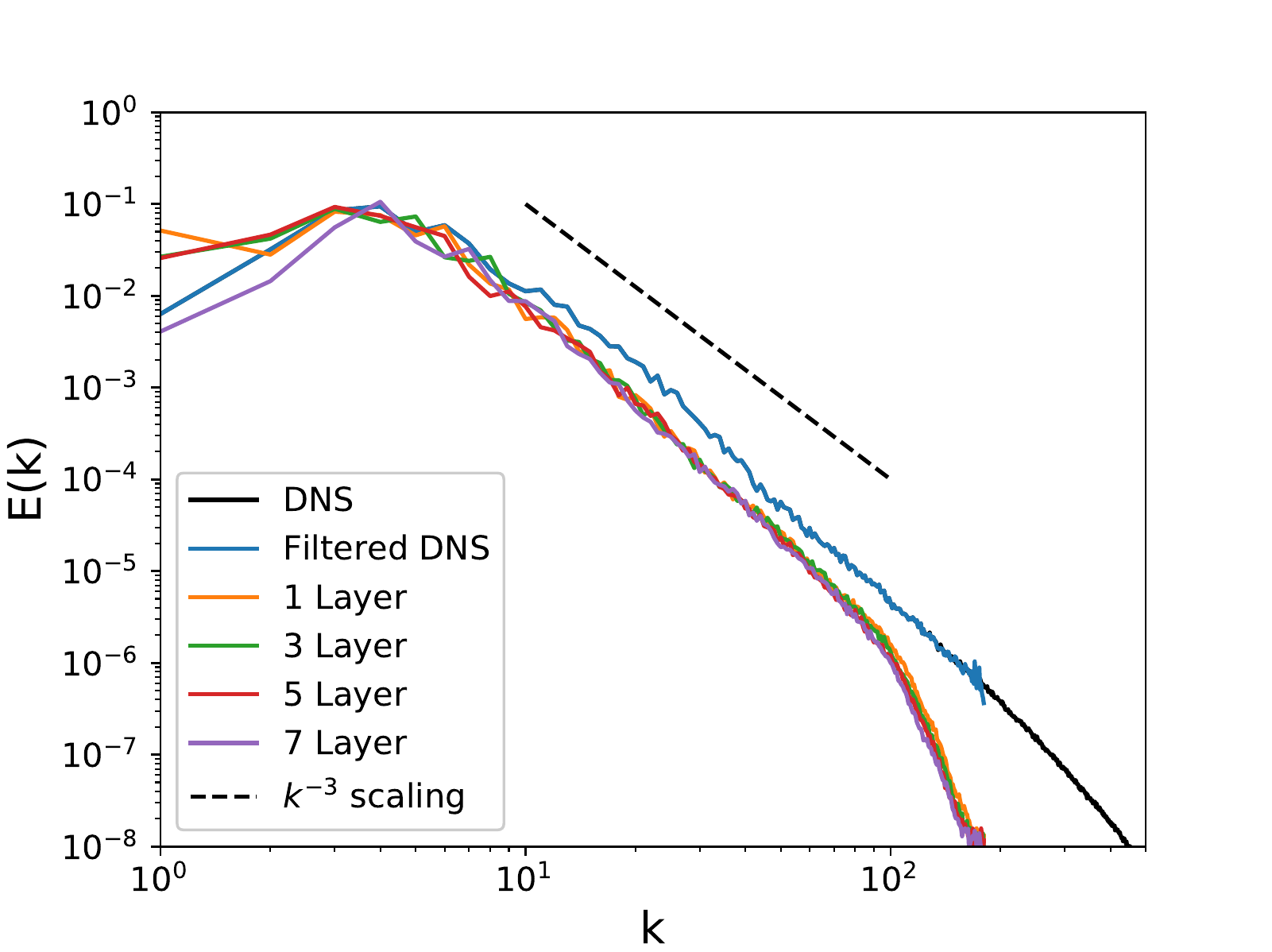}
\caption{Sensitivity study for proposed framework number of layers at $Re=64000$. Training data limited to $Re=32000$ only and with 100 neurons in each layer.}
\label{Fig12}
\end{figure}

\begin{figure}
\centering
\includegraphics[width=\columnwidth]{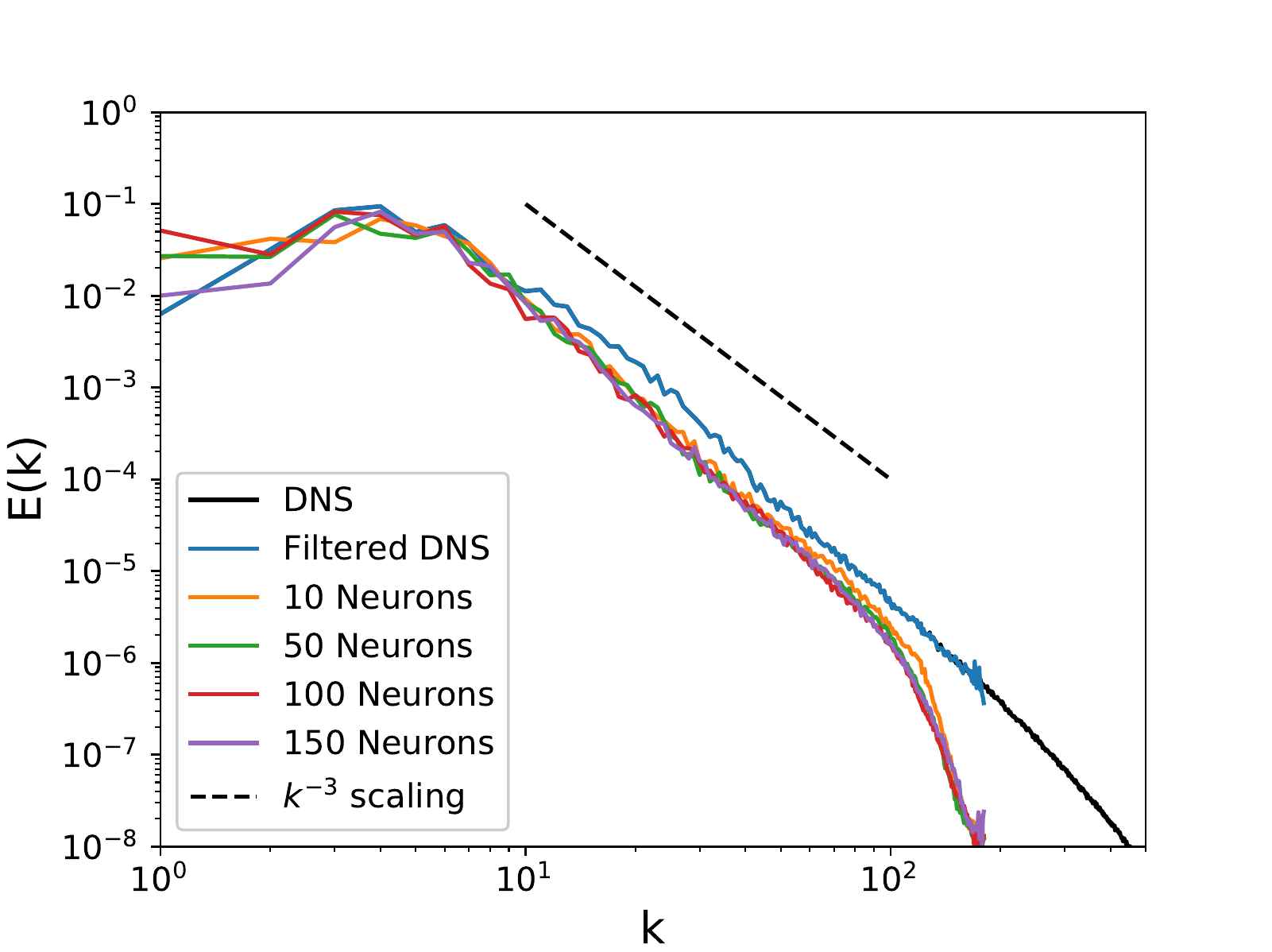}
\caption{Sensitivity study for proposed framework number of layers at $Re=64000$. Training data limited to $Re=32000$ only and with 1 hidden layer only.}
\label{Fig13}
\end{figure}

Figure \ref{Fig12} shows the effect of varying network depths, where it can be seen that a one-layered architecture performs sufficiently accurately to be considered optimal for deployment. This hints at a simpler nonlinear relationship between the inputs and outputs which has been captured by our framework. Figure \ref{Fig13} shows the effect of the number of neurons, where once again, it is observed that reduced model complexity does not impede performance. While this study utilized 100 neurons in the single hidden layer, even 10 would suffice for accurate scaling recovery. These observed behaviors imply that our framework allows for reduced network depths and reduced neurons and their associated computational advantages during training and deployment. However, we must caution the reader that greater amounts of data would necessitate deeper architectures for more generalization. In particular, our expectation is that if multiple flow scenarios were to be learned, simple feed-forward ANNs may prove to be inadequate. In particular, we note that our choice of localized sampling, network architecture and training loss-function are chosen specific to the resolution loss and physics at hand. Greater generalization (through improved diversity of training data) would require revised hyperparameter study.

\begin{figure}
\centering
\includegraphics[width=\columnwidth]{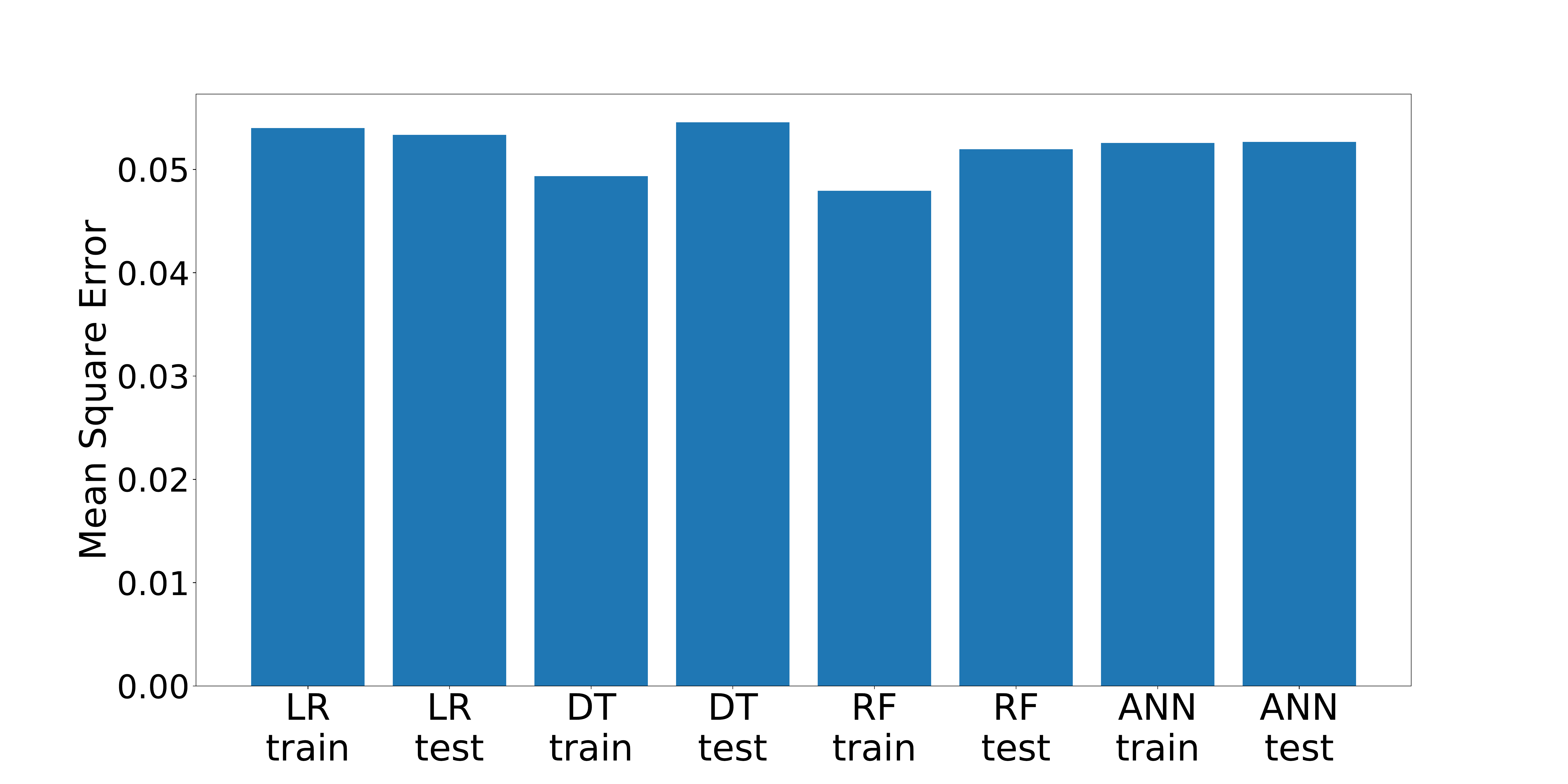}
\caption{Sensitivity study for machine learning algorithm for training and testing mean-squared-errors. These errors are shown for $\mathbb{M}_2$ optimization.}
\label{Fig14}
\end{figure}

For our problem of choice, it is evident that a 10 neuron, 1 layer ANN is sufficiently viable for estimating both $\mathbb{M}_1$ and $\mathbb{M}_2$. This lends evidence to the fact that our dual network formulation may also allow for simpler learning algorithms (i.e., for this particular problem). We perform an \emph{a priori} sensitivity study for training and test mean-squared-error measures for three other well-known statistical learning algorithms such as a linear regressor (LR), a random-forest regressor (RF) \cite{liaw2002classification} and a decision-tree regressor (DT) \cite{safavian1991survey}. We utilize the open-source scikit-learn machine learning library in python for standard implementations of these techniques. A quantitative training and testing mean-squared-error performance for these techniques in comparison to the ANN is shown in Figure \ref{Fig14} where it is observed that similar performance characteristics are observed despite vastly different learning methodologies for $\mathbb{M}_2$ optimization. It can thus be concluded that the utilization of our dual network framework has led to the simplification of a highly nonlinear problem to one that is tractable for linear learning methods.

\begin{figure}
\centering
\includegraphics[width=\columnwidth]{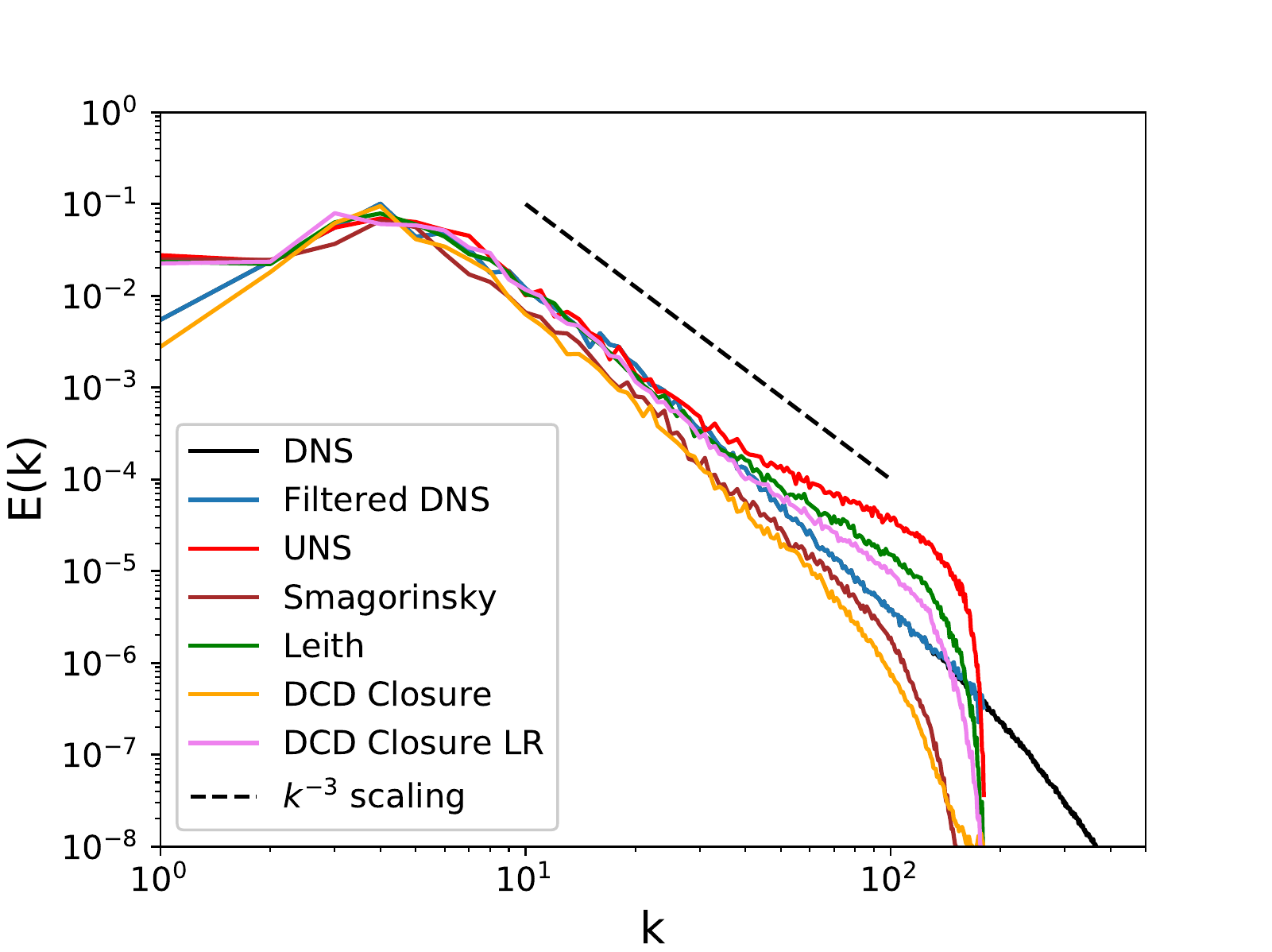}
\caption{The performance of a linear estimator (LR) for convolutional and deconvolutional maps in the proposed framework for $Re=32000$. A comparison to the default ANN is shown.}
\label{LinPos1}
\end{figure}

\begin{figure}
\centering
\includegraphics[width=\columnwidth]{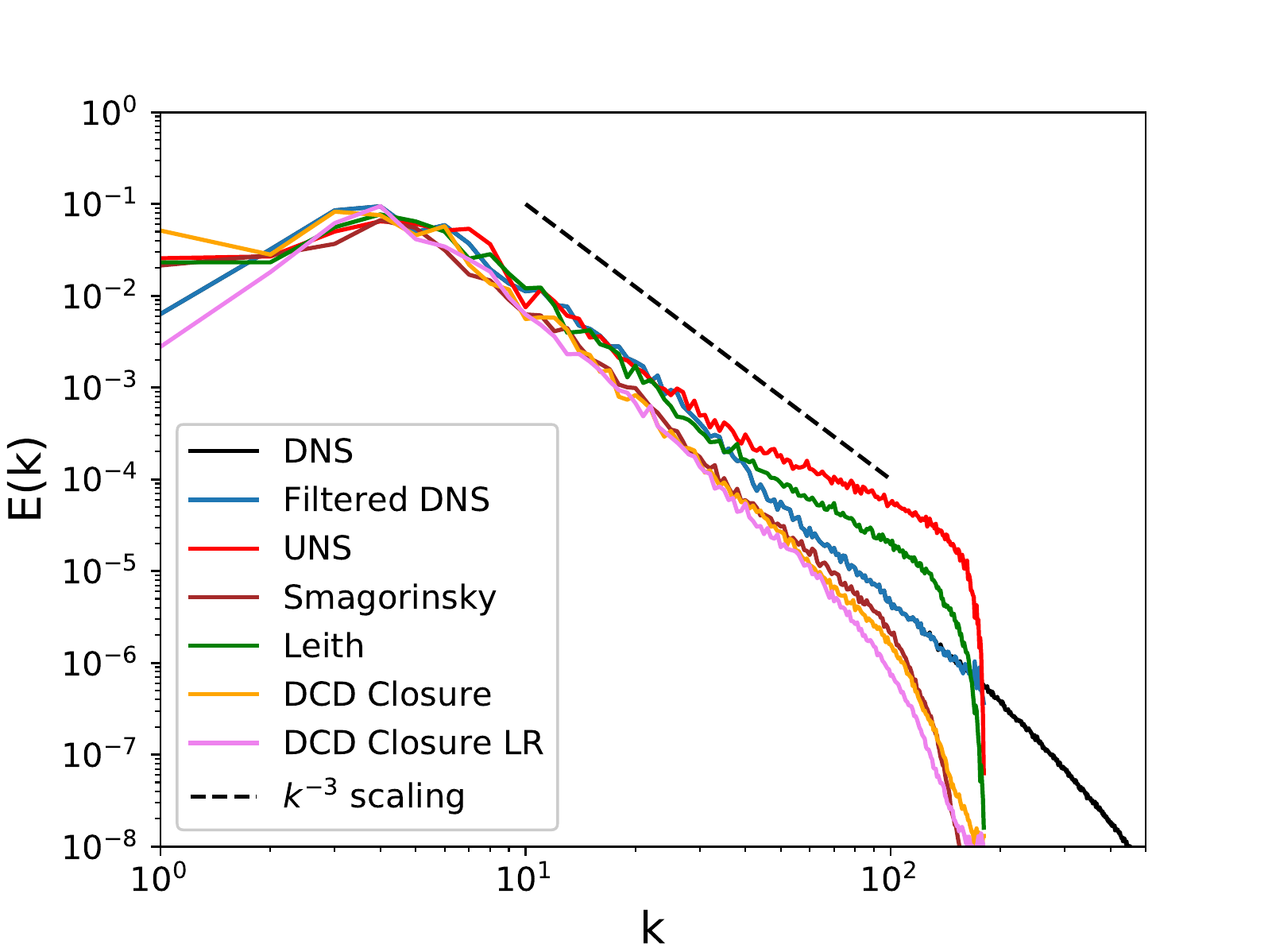}
\caption{The performance of a linear estimator (LR) for convolutional and deconvolutional maps in the proposed framework for $Re=64000$. A comparison to the default ANN is shown. Training data limited to $Re=32000$ only.}
\label{LinPos2}
\end{figure}

The linear-regressor is also implemented in an \emph{a posteriori} manner as shown in Figures \ref{LinPos1} and \ref{LinPos2} for $Re=32000$ and $Re=64000$ respectively. The kinetic energy spectra predictions of these linear relationships which estimate the convolutional and deconvolutional relationships are slightly less dissipative in the inertial and grid cut-off length scales for the $Re=32000$ case. However, very similar performance is obtained for $Re=64000$. The slightly discrepancy in the performance of the linear implementations of the convolutional and deconvolutional maps may be attributed to a lower generalizability of the simpler nature of its learning. However, we would like to remark that this has positive implications for the utility of these techniques for the preservation of the solenoidal constraint and frame-invariance in higher-dimensional flows \cite{stolz1999approximate} on structured grids. We would also like to note that the utilization of the same data-local filter stencil in all locations of the specified mesh ensures Galilean invariance \cite{razafindralandy2007analysis}. In addition, the use of stencil inputs is philosophically aligned with \cite{moser2009theoretically}, where multipoint input data are used for optimal LES formulations. However, further research is necessary for importing concepts related to isotropization of these data-driven filter and inverse kernels for application to general unstructured grids. It is also necessary to explore the possibilities of `constrained-learning' which may embed the preservation of the solenoidal constraint in higher-dimensions through penalties introduced to the loss-functions \cite{raissi2018hidden}. That is a subject of on-going investigation.

\section{Modified truncation via mean filtering}

\begin{figure}
\centering
\mbox{
\includegraphics[width=\columnwidth]{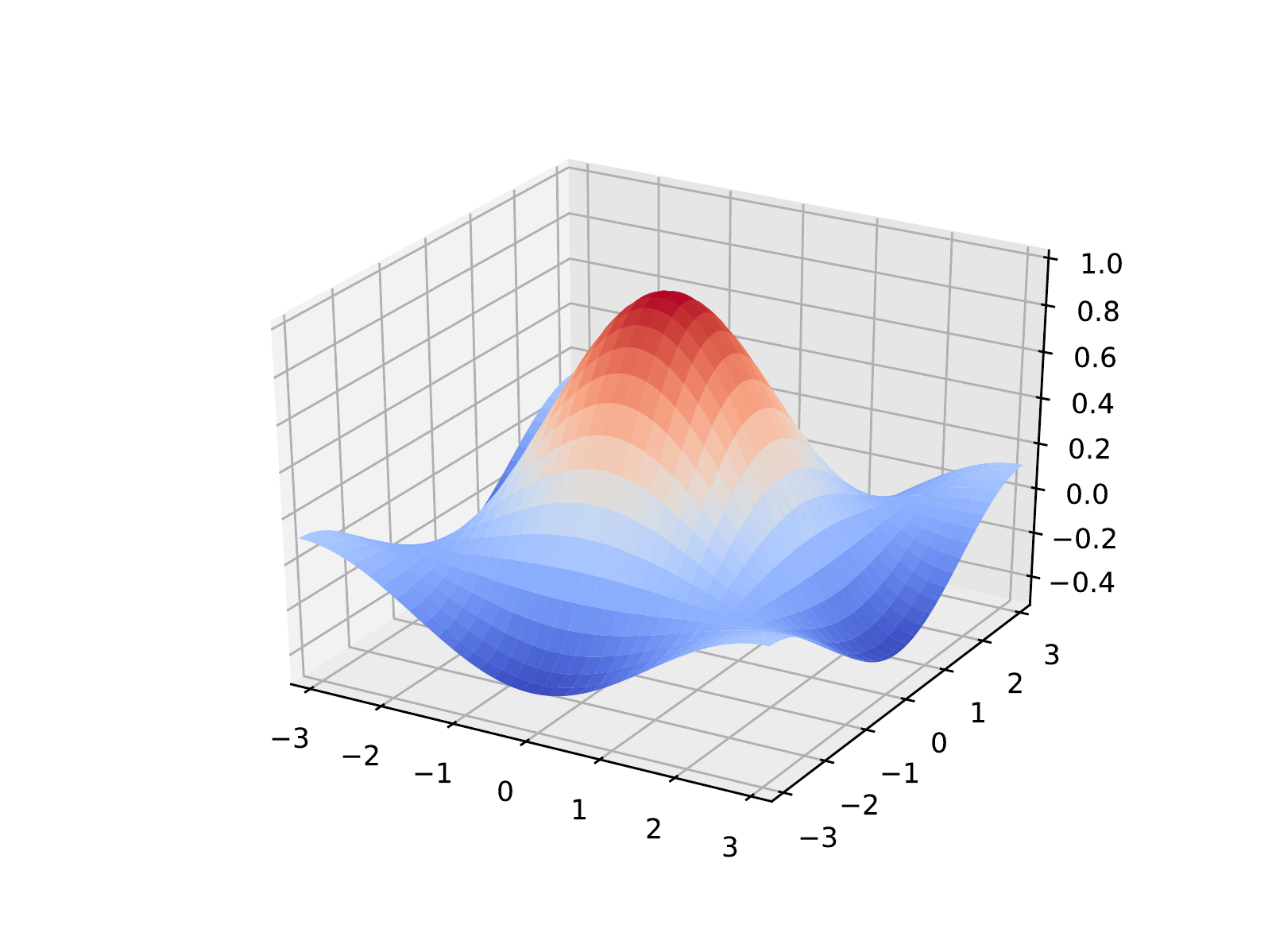}
}
\caption{Transfer function for truncation kernel to preserve statistical effects of backscatter.}
\label{FigBS}
\end{figure}

The truncation specified in Equation \ref{Eq12} and Figure \ref{Fig4} leads to an asymmetry in the estimation of the dissipation by finer wavenumbers. To that end, we introduce a modified truncation kernel based on a local-averaging for an added truncation of positive eddy-viscosity predictions to ensure a balance with backscatter. This is introduced through the concept of a locally-averaged eddy-viscosity prediction, for instance, given by
\begin{align}
\label{EqRev2}
\nu^{av}_{i,j} = \frac{1}{9}\left(\nu_{i,j}^e + \nu_{i,j+1}^e + \nu_{i,j-1}^e  + \hdots + \nu_{i-1,j-1}^e \right),
\end{align}
where
\begin{align}
\nu_{i,j}^e = \frac{\tilde{\Pi}_{i,j}}{\nabla^2 \bar{\omega}_{i,j}}.
\end{align}
The averaging procedure in Equation \ref{EqRev2} may also be represented by a mean-filtering-kernel given as
\begin{align}
\nu^{av} = \frac{\nu^e}{9}
\begin{bmatrix}
1 & 1 & 1 \\
1 & 1 & 1 \\
1 & 1 & 1
\end{bmatrix}.
\end{align}
The transfer-function of this kernel may be visualized as shown in Figure \ref{FigBS} and this averaging filter has the effect of eliminating localized pointwise values which are unrepresentative of their surroundings.

\begin{figure}
\centering
\includegraphics[width=\columnwidth]{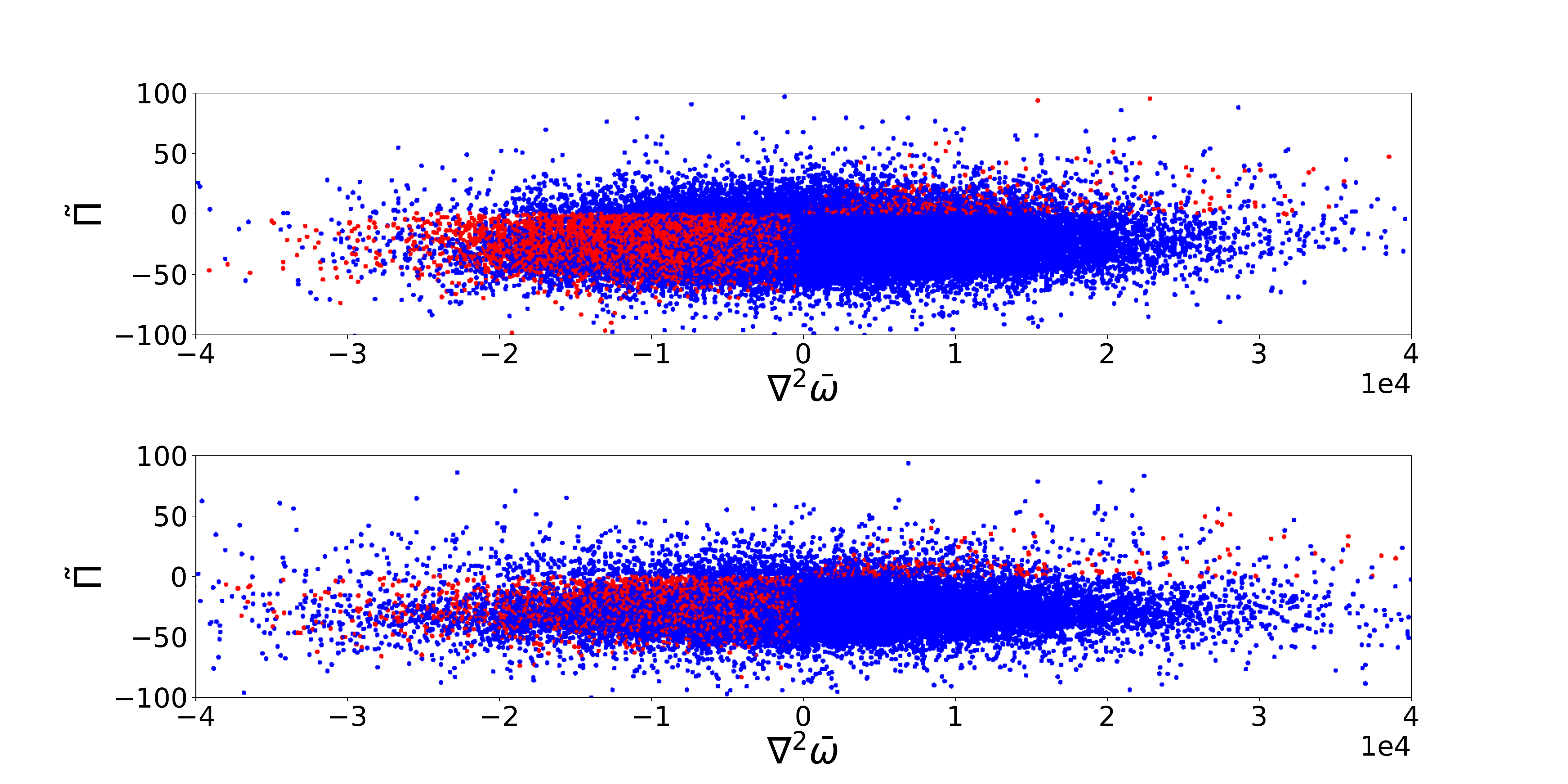}
\caption{A visual assessment of the truncation of our numerical post-processing during deployment given by the BS-1 framework. Blue points indicate truncated deployment for ensuring no negative viscosity and numerical stability. A-priori predictions for $Re=32000$ (top) and $Re=64000$ (bottom) shown. }
\label{FigRev5a}
\end{figure}

The quantity $\nu^{av}_{i,j}$ is basically the averaged dissipative (or energy-producing) nature of the local stencil of prediction and the quantity $\nu_{i,j}^e$ is the local effective eddy-viscosity prediction by our proposed framework. Our truncation scheme is then expressed as
\begin{align}
\label{EqRev3}
\Pi_{i,j} =
\begin{cases}
\tilde{\Pi}_{i,j},& \text{if  } \nu^{av}_{i,j} > \nu^{e}_{i,j}\\
    0,              & \text{otherwise.}
\end{cases}
\end{align}
The effect of this modified truncation is described in Figure \ref{FigRev5a} where an increased truncation is observed quite clearly. Our model formulation may thus be assumed to preserve the statistical nature of the negative-eddy viscosities in a locally-averaged manner.

% \begin{figure}
% \centering
% \includegraphics[width=\columnwidth]{Backscatter_Scatter_BS_2.pdf}
% \caption{A visual assessment of the truncation of our numerical post-processing during deployment given by the BS-2 framework. Blue points indicate truncated deployment for ensuring no negative viscosity and numerical stability. A-priori predictions for $Re=32000$ (top) and $Re=64000$ (bottom) shown. }
% \label{FigRev5b}
% \end{figure}

\begin{figure}
\centering
\includegraphics[width=\columnwidth]{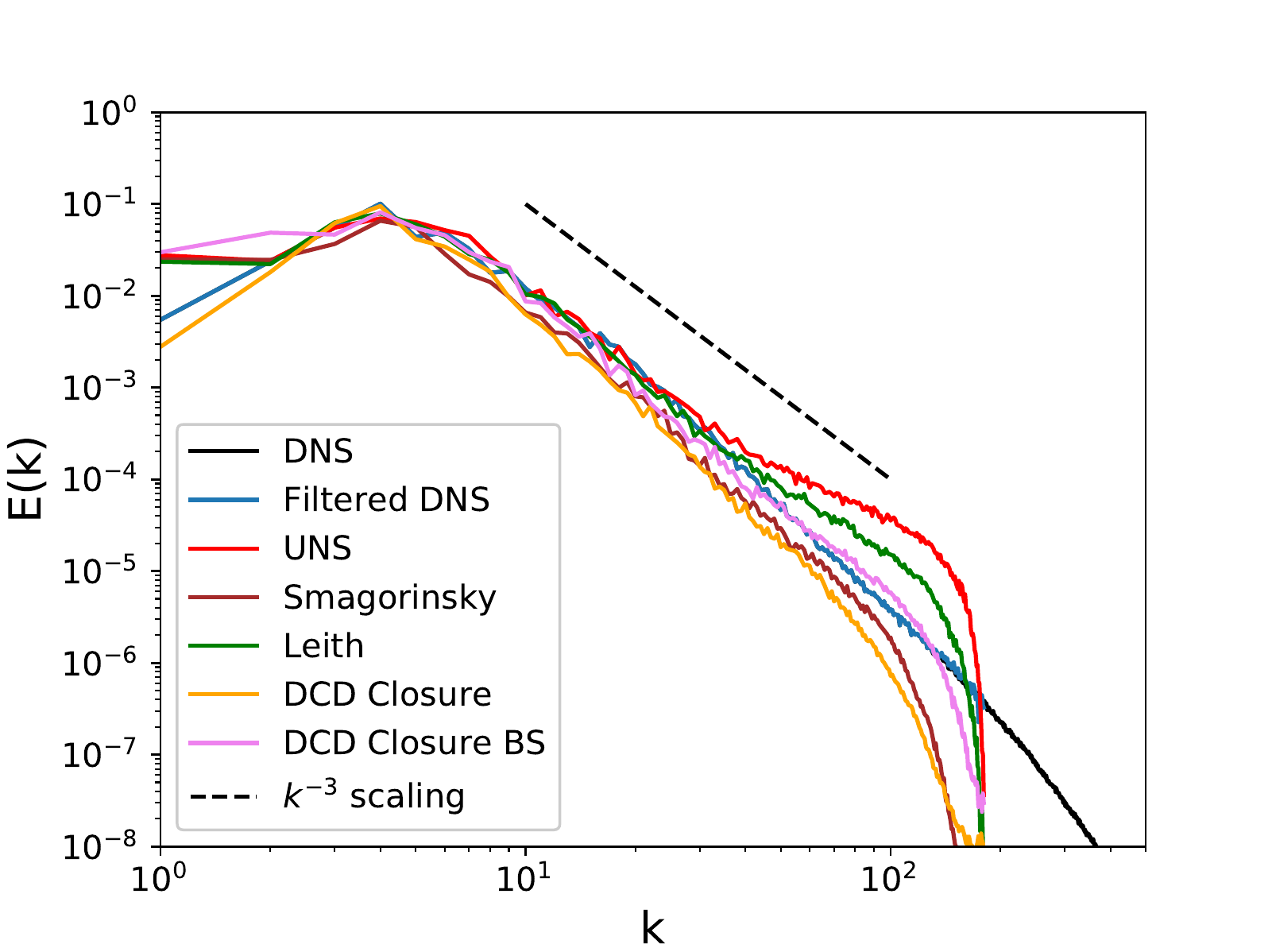}
\caption{A comparison of the choice of \emph{a posteriori} truncation utilized in our proposed framework. A statistical preservation of backscatter enforced by our proposed kernel leads to a better agreement with the inertial range statistics for $Re=32000$.}
\label{FigRev6}
\end{figure}

\begin{figure}
\centering
\includegraphics[width=\columnwidth]{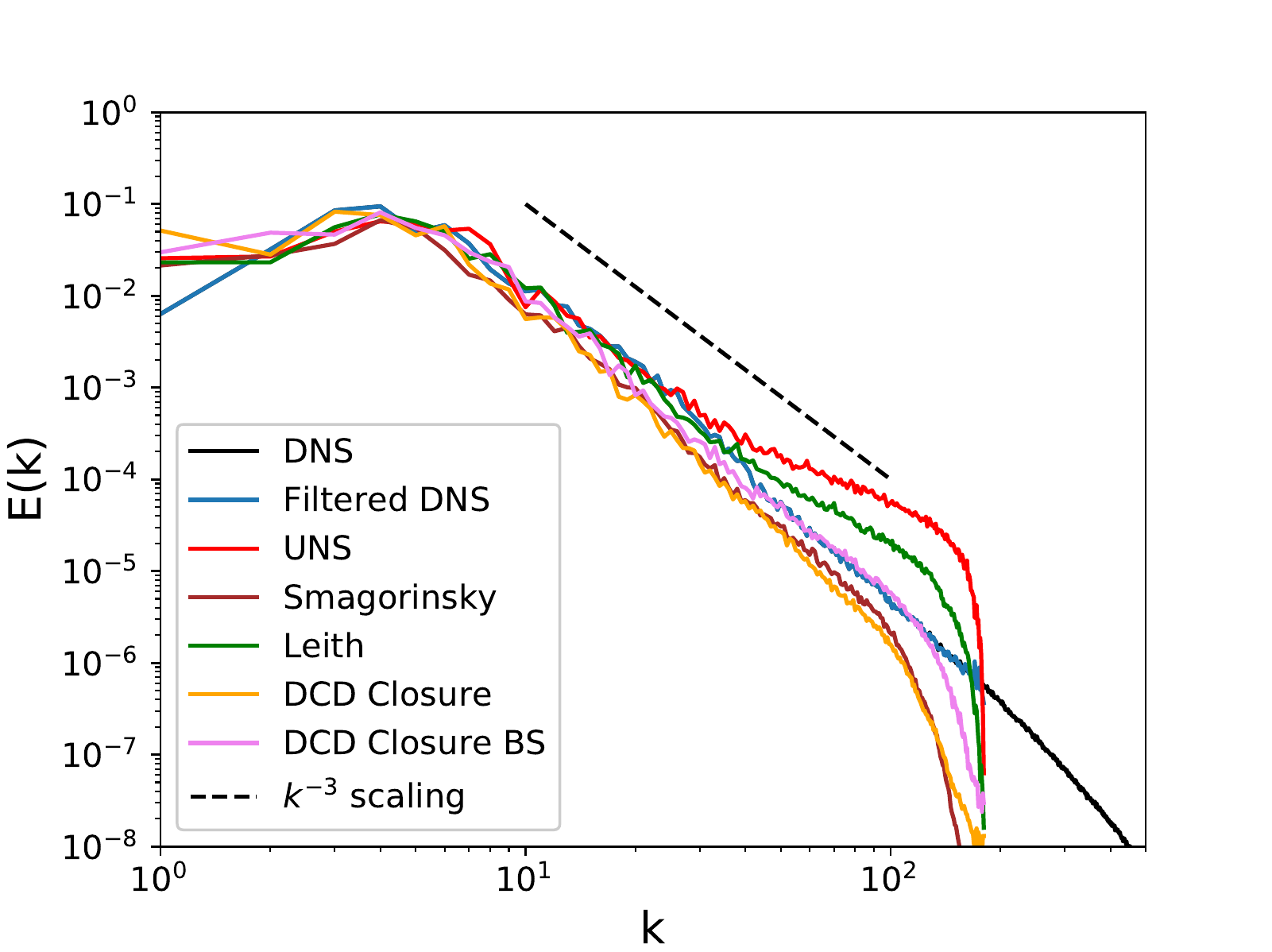}
\caption{A comparison of the choice of \emph{a posteriori} truncation utilized in our proposed framework. A statistical preservation of backscatter enforced by our proposed kernel leads to a better agreement with the inertial range statistics for $Re=64000$. Training data limited to $Re=32000$ only.}
\label{FigRev7}
\end{figure}

\emph{A posteriori} deployments of this modified truncation scheme are displayed in Figures \ref{FigRev6} and \ref{FigRev7} where an improved capture of the inertial range is observed for $Re=32000$ and $Re=64000$ respectively. This implies that the statistical fidelity of the prediction has been improved by the integration of a local backscatter estimate. The combination of novel truncation strategies may further be studied in the context of this data-driven framework for close agreement with theoretical scaling laws.

\section{Concluding remarks}

In this investigation, we have put forth and analyzed a physics-informed data-driven closure modeling framework for nonlinear partial differential equations. Our proposal is to use two single-layer feed-forward artificial neural networks for mapping transformations from grid-resolved variables with missing wavenumber content and subsampled direct numerical simulation data in order to close the two-dimensional Navier-Stokes equations. This investigation continues from the authors' previous work \cite{maulik2017neural}, which assessed the deconvolutional ability of neural networks, by employing them for estimating sub-grid relationships from grid-resolved variables.

Our framework precludes the utilization of any phenomenological arguments or model form constraints and relies, instead, solely on the approximation of the Fourier cut-off filtering inherent in coarse-graining as well as its approximate inverse. We remark that while there is truly no way to invert a Fourier cut-off filter, a-priori exposure to samples from resolved and filtered fields are used to estimate the information loss and reconstruct it. For the purpose of numerical stability, we also employ two postprocessing strategies with the first ensuring no aggregate negative viscosities in the computational domain and the second preserving backscatter in a statistical sense. This ensures that the stochastic nature of the network predictions do not trigger numerical instability amplification in an explicit flow computation. Of, the two proposed truncation mechanisms for the preservation of backscatter, our first formulation shows a good agreement with DNS statistics whereas the second truncates excessively. However, we note that the many such kernels may be investigated and we seek to undertake this for future research.

Another important feature of this investigation is that, despite its data-driven nature, our offline training phase necessitates no exposure to the true sub-grid stress data and predictions are viable simply through the estimation of the nature of the coarse-graining process in LES. Our sensitivity study reveals the benefits of this approach, where it is seen that increasing network complexity leads to no appreciable improvement in the \emph{a posteriori} performance for this current test case. The need for complicated network architectures (and their associated computational and memory burden) is thus minimized due to the physics-informed nature of our formulation.

Comparison with other well-established linear statistical learning methods also show that the novel dual network formulation presented here reduces the complexity of learning considerably. In particular, the performance of a linear map representation of convolution and deconvolution operations ensures a direct enforcement of the solenoidal constraint on the convolved and deconvolved fields for applicability to higher dimensions. \emph{A posteriori} realizations of the linear mappings between grid-resolved and sub-grid space, show an exhibition of the bias-variance trade-off issue where the simpler nature of the linear regressor leads to lower generalization for a different data-set. However, an effective parameter and model-form free closure is readily obtained in this case as well.

We also note that the data-local nature of our framework with the combination of solely one map (each for convolution and deconvolution) ensures that frame-invariance is respected for the specified mesh. As a future direction, this framework shall be studied with the view of integrating physics-based constraints in the offline training phase. These may be introduced through optimization penalties for continuity enforcement and for isotropization on arbitrary meshes. These are necessary for the generalization of this framework to higher-dimensional flows with arbitrary boundary conditions.

While the results of this study have proven promising for the development of purely data-driven closures for LES, the true test of these ideologies would be to develop generalized closures for a variety of flows. In terms of a long-term goal, the preliminary results displayed here must translate to a situation where \emph{a posteriori} closure is determined by \emph{a priori} exposure to a variety of flow classes. Additionally, the stencil based formulation for a predictive map leads to a resolution dependence of the trained relationships. This is because our LES to DNS ratio is fixed during the specification of training data. An exposure to different levels of coarse-graining for potential predictions would also increase the generalizability of this framework. With that in mind, we remark that the framework proposed here represents the advantages of implementing a data-driven paradigm from a physics-informed point of view with consequent benefits for framework complexity and ease of deployment.

\begin{acknowledgements}
This material is based upon work supported by the U.S. Department of Energy, Office of Science, Office of Advanced Scientific Computing Research under Award Number DE-SC0019290. O.S. gratefully acknowledges their support. Direct numerical simulations for this project were performed using resources of the Oklahoma State University High Performance Computing Center. Disclaimer: This report was prepared as an account of work sponsored by an agency of the United
States Government. Neither the United States Government nor any agency thereof, nor any of their
employees, makes any warranty, express or implied, or assumes any legal liability or responsibility
for the accuracy, completeness, or usefulness of any information, apparatus, product, or process
disclosed, or represents that its use would not infringe privately owned rights. Reference herein to
any specific commercial product, process, or service by trade name, trademark, manufacturer, or
otherwise does not necessarily constitute or imply its endorsement, recommendation, or favoring by
the United States Government or any agency thereof. The views and opinions of authors expressed
herein do not necessarily state or reflect those of the United States Government or any agency thereof.
\end{acknowledgements}

\bibliography{references}
\end{document}